\documentclass[a4,aps,twocolumn,showpacs,preprintnumbers,qamslatex,amsmath,amssymb,longbibliography]{revtex4-1}
\usepackage[cp1251]{inputenc}
\usepackage[T2A]{fontenc}
\usepackage{fancyhdr}
\usepackage[colorlinks]{hyperref}
\usepackage{graphicx}
\usepackage{verbatim}
\usepackage{indentfirst}
\usepackage{latexsym}
\usepackage{dsfont}
\usepackage{epstopdf}

\begin{document}

\title{Bilinear-biquadratic Spin-1 rings:\\ an SU(2)-symmetric MPS algorithm for periodic boundary conditions}
%\date{\today}
\author{Mykhailo V. Rakov$^1$ and Michael Weyrauch$^2$}
\affiliation{$^1$ Kyiv National Taras Shevchenko University, 64/13 Volodymyrska Street, Kyiv 01601, Ukraine}
\affiliation{$^2$ Physikalisch-Technische Bundesanstalt, Bundesallee 100, D-38116 Braunschweig, Germany}

\begin{abstract}
An efficient algorithm for SU(2) symmetric matrix product states (MPS) with periodic boundary conditions (PBC) is proposed and implemented. It is applied to a study of the spectrum and correlation properties of the spin-1 bilinear-biquadratic Heisenberg model. We  characterize the various phases of this model by  the lowest states of the spectrum with angular momentum $J=0,1,2$ for systems of up to 100 spins. Furthermore, we provide precision results for the dimerization correlator as well as the string correlator.
\end{abstract}


\maketitle

\section{Introduction}

About 25 years ago the density-matrix renormalization group (DMRG) emerged as a
precision tool for the numerical description of one-dimensional quantum many-body systems~\cite{PhysRevB.48.10345}.
Starting with Ref.~\cite{PhysRevB.48.3844} it was applied successfully to many spin and strongly interacting electron systems.
Later it was realized that DMRG can be reformulated in terms of matrix product states (MPS)~\cite{PhysRevLett.75.3537, PhysRevB.55.2164}.
A comprehensive review of MPS algorithms and their relation to DMRG is presented in Ref.~\cite{Schollwoeck2011}.
In this review most
algorithms are formulated without regard to symmetries such as U(1) or SU(2).

However, already {\"O}stlund and Rommer~\cite{PhysRevLett.75.3537,PhysRevB.55.2164} in an attempt to understand the success of DMRG
used SU(2) symmetric MPS. They realized that
the local matrices (tensors) decompose into a structural part given by the symmetry and a degeneracy part. The structural part consists of a Clebsch-Gordan coefficient, and the (variational) parameters of the model
reside in the degeneracy part only. The number of parameters to be determined is therefore significantly
reduced with respect to a non-symmetric theory. This approach, first suggested by {\"O}stlund and Rommer, was  generalized to
higher order tensors in Ref.~\cite{PhysRevA.82.050301} and then applied
to MERA tensor network calculations~\cite{PhysRevB.83.115125, Singh2010, PhysRevB.86.195114}.

However, in addition to this basic implementation of symmetry into a tensor network with the purpose of reducing the number of independent parameters,
it is possible to entirely eliminate the structural tensors and develop algorithms in terms of the degeneracy tensors only~\cite{Dukelsky1998,McCulloch2001,McCulloch2002,McCulloch2007}.  This reduces the requirements for  computational resources significantly, and in turn enables significant improvements to the accuracy of the results that can be obtained.
The elimination of structural tensors from the algorithms requires  changes to standard (non-symmetric) tensor network implementations, e.g. via precomputation schemes as suggested in Ref.~\cite{PhysRevB.86.195114}.

It is the purpose of the present paper to develop an algorithm for SU(2) symmetric MPS with periodic boundary conditions (PBC) in terms of reduced tensors only.
We stress that we express symmetric tensors (using the  Wigner-Eckart theorem) in terms of reduced tensors (degeneracy tensors) and structural tensors consisting of products of Clebsch-Gordan coefficients. To this end we strictly follow the methods and conventions of Edmonds~\cite{edmonds}. We do {\it not} use the
{tree decompositions} advocated in Ref.~\cite{PhysRevB.86.195114} for the representation of symmetric tensors. The linear maps  relating different tree decompositions of symmetric tensors derived in Ref.~\cite{PhysRevB.86.195114} directly correspond to expressions relating different {coupling schemes}~\cite{edmonds} in terms of Racah $6j$, Wigner $9j$, or more general symbols. Consequently, the precomputations for the linear maps suggested in Ref.~\cite{PhysRevB.86.195114} may be expressed in terms of such symbols.

Furthermore, we apply the proposed algorithm to a physically rather complex one-dimensional model: the spin-1 bilinear-biquadratic Heisenberg (BBH) model on a ring,
\begin{equation}\label{eq-bilbiq}
H=\sum_{i=1}^N \left[\cos\theta \, \vec s_i \otimes \vec s_{i+1}+\sin\theta \, (\vec s_i \otimes \vec s_{i+1})^2\right]
\end{equation}
with $N+1$ set to $1$ and $\vec s_i$ the spin-$s$ SU(2) matrix representations. In doing so we will reproduce a number of well-known results, e.g. a precise calculation of the Haldane gap in order to check the capabilities of the proposed algorithm. Moreover, we will study the spectrum of the BBH
Hamiltonian for PBC and various $\theta$, which have not yet been addressed in the literature. This will shed some light on old questions concerning the phase structure to be discussed below.

The BBH model describes the behavior of atomic spinor condensates in optical lattices~\cite{Orzel2001,PhysRevLett.95.240404,Rossini2006,PhysRevB.73.014410,PhysRevLett.93.250405} under certain conditions.
It also models the physics of some quasi-one-dimensional crystals, e.g., in LiVGe$_2$O$_6$~\cite{PhysRevLett.83.4176}
or Ni(C$_2$H$_8$N$_2$)$_2$NO$_2$ClO$_4$ (NENP)~\cite{0295-5075-3-8-013}. Furthermore, it was extensively used
as a test bed for new tensor network algorithms~\cite{PhysRevB.85.035130,PhysRevB.88.075133, PhysRevB.90.125154}. However, these algorithms do not explicitly implement SU(2) symmetry, and the methods developed here (and for infinite systems in Ref.~\cite{1742-5468-2016-8-083101}) could serve well to implement
SU(2) symmetry into those approaches.

The continuous SU(2) symmetry of the BBH model cannot be broken due to the Mermin-Wagner-Coleman theorem~\cite{PhysRevLett.17.1133,Coleman1973}. As a consequence the eigenstates of this Hamiltonian can be characterized by the total angular momentum quantum number $J$.
Due to the (discrete) translational symmetry, the eigenstates can also be labelled by the quasi-momentum quantum number $p$. However, unlike SU(2) symmetry, translational symmetry will not be built into our MPS ansatz explicitly in the present  paper. The BBH model has other symmetries not explicitly built into the MPS here, e.g. SU(3) symmetry at $\theta=\frac{1}{4}\pi$ and $\theta=-\frac{3}{4}\pi$, a symmetry not easily uncovered in the spin representation of the model~\cite{PhysRevB.65.180402}.

\begin{figure}
\unitlength1cm
\begin{picture}(7,6)(0,0)
 \put(0.5,0)  {\includegraphics[width=6cm]{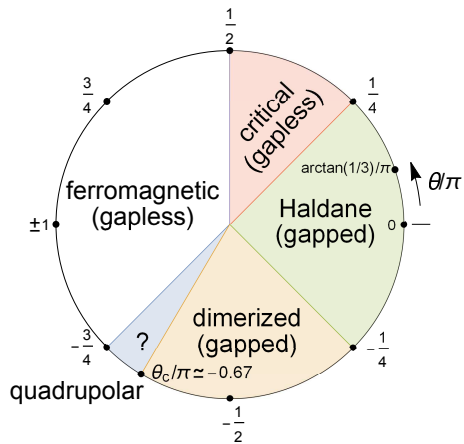}}
\end{picture}
\caption{\footnotesize (color online) Phase diagram of the infinite-size bilinear-biquadratic spin-1
Heisenberg model. The existence of the separate (nematic) phase at $-0.75 < \theta/\pi \lesssim -0.67$ has been long under debate.
\label{phased}}
\end{figure}

As a function of the control parameter $\theta$ the infinite-size spin-1 bilinear-biquadratic Heisenberg model exhibits the rich phase structure shown in Fig.~\ref{phased}~\cite{PhysRevB.84.054451}. We briefly discuss the various phases moving around the circle of the phase diagram in a clockwise direction.

For $\frac{\pi}{4}>\theta>-\frac{\pi}{4}$ the system is in the gapped Haldane phase with hidden topological order~\cite{Haldane1983,PhysRevB.81.064439}.
%Correlations decay exponentially in this phase.
The ground state has spin $J=0$ and is non-degenerate. The first excited state is a triplet (spin-1). At $\theta=0$ the model corresponds to the simple Heisenberg antiferromagnet and at $\theta=\arctan \frac{1}{3}$ the ground state is the AKLT state~\cite{PhysRevLett.59.799}. % with energy $E_0/N=-\frac{2}{3}$.
At $\theta=-\frac{\pi}{4}$ the gap vanishes and there is a second order phase transition to the dimerized phase~\cite{Takhtajan1982,Babujian1982}.
%At the critical point the ground state energy is given by $E_0/N=-\sqrt{8}$~\cite{Takhtajan1982}.

The dimerized phase extends from $\theta=-\frac{\pi}{4}$ to $\theta=-\frac{3\pi}{4}$. The ground state is a {\it doubly degenerate} $J=0$ state with a gap to either spin-1 or spin-2 excited state~\cite{Nomura1991}. It shows non-zero dimer order~\cite{PhysRevB.47.872,PhysRevB.40.4621}. At $\theta=-\frac{\pi}{2}$ the (biquadratic) model can be mapped to the spin-1/2 XXZ model and, therefore, it is exactly solvable using the Bethe Ansatz~\cite{PhysRevB.40.4621,PhysRevB.42.754}. This point is characterized by very long-ranged spin correlations~\cite{PhysRevB.42.754} and `maximum' nearest-neighbor entanglement.

According to~\cite{Kawashima2002,PhysRevB.74.144426}, the dimer order parameter behaves non-typically in the parameter region $\theta_c\approx-0.67\pi>\theta >-\frac{3\pi}{4}$, and quadrupolar spin correlations increase dramatically there. The conjecture of Chubukov~\cite{PhysRevB.43.3337} that there is a gapped nematic phase in this parameter region was debated in Refs.~\cite{PhysRevB.51.3620,PhysRevLett.95.240404} and rejected in~ Ref.~\cite{PhysRevLett.98.247202}. However, the existence of the nematic gapless phase in the narrow vicinity of $\theta=-3\pi/4$ is not ruled out, since the values to be calculated are comparable with the precision of the numerical calculations in that parameter region~\cite{PhysRevB.72.054433,PhysRevB.74.144426}.

In the region $\frac{5\pi}{4}>\theta>\frac{\pi}{2}$ the system is in a gapless ferromagnetic phase with a multi-fold degenerate ground state.
%The energies at the transition points $5\pi/4$ and $\pi/2$ are $E_0/N=-\sqrt{2}$ and $E_0/N=1$, respectively.
At the critical point $\theta=-\frac{3\pi}{4}$ the system exhibits SU(3) symmetry~\cite{PhysRevB.65.180402}.
The ferromagnetic phase is followed by a gapless phase with dominating quadrupolar spin correlations~\cite{PhysRevB.74.144426,PhysRevB.58.5498} for $\frac{\pi}{2}>\theta>\frac{\pi}{4}$.
At the Lai-Sutherland point $\theta=\frac{\pi}{4}$ the system undergoes a Kosterlitz-Thouless phase transition  ~\cite{PhysRevB.55.8295,PhysRevB.47.872} into the Haldane phase. At the Lai-Sutherland critical point the system shows SU(3) symmetry and is exactly solvable by the Bethe Ansatz (see, e.g.,~\cite{PhysRevB.12.3795, Nomura1991}).

Using the proposed SU(2) symmetric MPS algorithm for PBC, we calculate energy spectra and characteristic correlation
functions in different phases of a spin ring at selected  parameter values $\theta$. The calculated energy gaps between these states enable already an elucidation of the phase structure from finite system results. In particular, the existence of a fifth (nematic) phase can be addressed. For the Haldane phase we calculate the string correlator of the ground state and in the dimerized phase the dimerization  from the lowest two $J=0$ states. These two states form a degenerate doublet \textit{in the thermodynamic limit}.

The paper is organized as follows. In Sec.~\ref{standard} we briefly review the MPS formalism for periodic boundary conditions which is based on the algorithm proposed by
Verstraete, Porras, and Cirac~\cite{PhysRevLett.93.227205}.  In Sec.~\ref{SU2mps} this algorithm is rewritten
using SU(2) symmetric tensors only. As already emphasized, it is a major objective of the present paper
to eliminate all structural tensors from the PBC algorithm. This is more complicated than for OBC since higher order tensors must be considered. These technical developments are relegated to several appendices, which form an important part of the present article and should enable a straightforward implementation of the algorithm.

As a consequence we only need to handle the degeneracy parts of the tensors explicitly. This enables significant improvements to the efficiency of the implementation.
In fact, the MPS virtual dimensions we are able to use are much larger than in various recently
proposed implementations for PBC without SU(2) symmetry, e.g.~\cite{PhysRevB.85.035130,PhysRevB.88.075133}.
We provide the reduced MPO representation for the bilinear-biquadratic spin-$s$ Hamiltonian as well as the corresponding reduced representation for $H^2$ and other operators. This enables the calculation of the variance $\langle H^2\rangle -\langle H\rangle^2$ of the various eigen-energies.

We apply the proposed SU(2) symmetric algorithm to the bilinear-biquadratic spin-1 Heisenberg model in Sec.~\ref{applications} and discuss the low lying spectrum in all phases except the ferromagnetic phase.
We also briefly address in Appendix~\ref{spin-1/2} the spin-1/2 Heisenberg model and compare our numerical results with Bethe Ansatz calculations. The results of our work are summarized in section~\ref{conclusions}.

\section{Review: the MPS formalism for PBC}\label{standard}

An MPS formalism for PBC was originally proposed in~\cite{PhysRevLett.93.227205} and extended in~\cite{1367-2630-14-12-125015} and~\cite{PhysRevB.73.014410}. We summarized the algorithm in Refs.~\cite{Weyrauch2013, PhysRevB.93.054417}, and therefore we only briefly review here those aspects  which are relevant for the present discussion.

The state of a 1D quantum many body system of size $N$ is approximated in terms of a matrix product
state
\begin{equation}\label{eq:MPS}
|\psi\rangle=\sum_{ \{\sigma\}, \{a\}} M_{a_0,a_1}^{[1],\sigma_1} \, \cdots
\, M_{a_{N-1},a_N}^{[N],\sigma_N} |\sigma_1 \dots \sigma_N\rangle.
\end{equation}
Here, the $\sigma_i$ represent the local degrees of freedom at site
$[i]$. The local Hilbert space is assumed to be finite dimensional, and its basis is enumerated by the $\{\sigma_i\}$. The elements $M_{a_{i-1},a_i}^{[i],\sigma_i}$ of the rank-3 tensors $M^{[i]}$ are the parameters characterizing a particular state.  For periodic systems we set $a_0=a_N$, and the dimension of the bond indices $a_i$ is set to $m$.

Analogously, operators are written as matrix product operators
\begin{eqnarray}\label{eq-MPO}
O&=&\sum_{\{\sigma\},\{\sigma^{\prime}\}, \{I\}} \, W_{I_0,I_1}^{[1],\sigma_1,\sigma_1^\prime} \, \dots \, W_{I_{N-1},I_N}^{[N],\sigma_N,\sigma_N^\prime} \\
& &~~~~~~~~~~~~~~~~~~~~~~~~~~~~~~~|\sigma_1 \dots \sigma_N \rangle\langle \sigma_1^\prime \dots \sigma_N^\prime|.\nonumber
\end{eqnarray}
Each $W^{[i]}$ is a rank-4 tensor: $W_{I_{i-1},I_i}^{[i],\sigma_i,\sigma_i^{\prime}}$ with $I_0=I_N$. The bond indices $I$ of operators have dimension $m_W$, and the operators studied in
the present work can be
expressed {\it exactly} in terms of $W$ tensors with small bond dimension.

Matrix elements of MPO in MPS,
\begin{equation}\label{eq:matrixe}
\langle\phi|O|\psi\rangle={\rm Tr}\; (E_W^{[1]}(A,B)\cdot\ldots \cdot E_W^{[N]}(A,B)),
\end{equation}
can be expressed in terms of the rank-6 (generalized) transfer tensor $E_W^{[i]}$ with the tensor elements
\begin{equation}\label{EWW}
(E_W^{[i]})_{(\bar{I},\bar{a},\bar{b}),(I,a,b)}
=\sum_{\sigma,\sigma^\prime} \, W_{\bar{I},I}^{[i],\sigma,\sigma^\prime} \, A_{\bar{a},a}^{*[i],\sigma} \, B_{\bar{b},b}^{[i],\sigma^\prime}.
\end{equation}
The matrices $A$ and $B$ characterize the states $|\phi\rangle$ and
$|\psi\rangle$, respectively. The special
transfer tensor $E_1^{[i]}(A,B)$ represents the matrix element
$\langle \phi | \psi \rangle$ of the identity operator (with $\bar{I}=I=1$). Note, that we chose to write
the algorithm in terms of higher-rank tensors in order to ease the implementation of SU(2) symmetry in the next section. Parentheses are used to group indices conveniently.
\begin{figure}
\unitlength1cm
\begin{picture}(6,3)(0,0)
 \put(-0.25,0)  {\includegraphics[width=6cm]{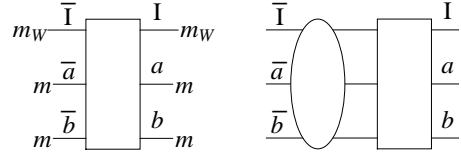}}
% \put(3,0)  {\includegraphics[width=3.5cm]{update.eps}}
\end{picture}
\caption{\footnotesize (left) Graphical representation of the transfer tensor $E_W$ Eq.~(\ref{EWW}).
(right) Enlarging a `block' tensor (oval) by one transfer tensor
(rectangle) on the right. The lines connecting the block tensor and transfer tensor represent
contractions.
\label{fig:tm}}
\end{figure}

In order no minimize the energy variationally, due to PBC  one has to solve at each update step the {\it generalized} eigenvalue problem
\begin{equation}\label{eq:eigenproblem-p}
H_{\rm eff}^{[i]} \, \nu^{[i]}=\epsilon^{[i]} \, N_{\rm eff}^{[i]} \, \nu^{[i]}
\end{equation}
written in terms of the effective Hamiltonian matrix $H_{\rm eff}$ and the effective normalization matrix $N_{\rm eff}$. The elements of these matrices are obtained from
\begin{eqnarray}\label{eq:eff-Hamiltonian-p}
&&(H_{\rm eff}^{[i]})_{[\sigma_i a_i a_{i-1}],[\sigma_i^{\prime} a_i^{\prime} a_{i-1}^{\prime}]}=%\sum_{a_N,a_N^{\prime}}
\sum_{k l}^{m_W} W_{k,l}^{[i],\sigma_i,\sigma_i^{\prime}} \times \\
&&~~~~ (H_R^{[i]})_{(l, a_i, a_i^{\prime}),[p a_N a_N^{\prime}]} \cdot (H_L^{[i]})_{[p a_N a_N^{\prime}],(k, a_{i-1}, a_{i-1}^{\prime})},\nonumber
\end{eqnarray}
\begin{eqnarray}\label{eq:eff-norm-p}
&&(N_{\rm eff}^{[i]})_{[\sigma_i a_i a_{i-1}],[\sigma_i^{\prime} a_i^{\prime} a_{i-1}^{\prime}]}= %\sum_{a_N,a_N^{\prime}}
\delta_{\sigma_i \sigma_i^{\prime}} \times \\
&&~~~~(N_R^{[i]})_{(1, a_i, a_i^{\prime}),[1 a_N a_N^{\prime}]} \cdot (N_L^{[i]})_{[1 a_N a_N^{\prime}],(1, a_{i-1}, a_{i-1}^{\prime})}.\nonumber
\end{eqnarray}
The `blocks' $H_L^{[i]},~N_L^{[i]}$ and $H_R^{[i]},~N_R^{[i]}$ are contractions of the transfer tensors $E_W$ and $E_1$ (as defined in Eq.~(\ref{EWW})) to the left or to
the right of the site $i$, respectively. (In order to define what is left and right we initially label the sites from 1 to $N$, and keep this labeling throughout the calculation. All sites with label smaller than $i$ are left of site $i$ and all sites with label larger than $i$ are right of site $i$.) A contraction of a block with a transfer tensor is depicted in Fig.~\ref{fig:tm}.
The block  and transfer tensors are kept rank-6 throughout the algorithm. We assume summation over repeated indices,
however, for clarity we sometimes write out the summations explicitly. Brackets are used in order to indicate
fusion of several indices into a combined single index.
The update is performed in the standard DMRG `sweeping' manner as proposed in Ref.~\cite{PhysRevLett.93.227205}.

The value $\epsilon^{[i]}$ eventually converges to the ground state energy during the iterative update procedure. The updated MPS is obtained by a suitable partitioning of the vector into a tensor:
\begin{equation}\label{updMPS}
\nu_{[\sigma_i a_i a_{i-1}]}^{[i]}=M_{a_{i-1},a_i}^{[i],\sigma_i}.
\end{equation}
After each update step the local MPS tensor is regauged in order to keep the algorithm stable, i.e.
we modify each local MPS tensor $ M^{[i]}$ such that one of the following relations is fulfilled,
\begin{eqnarray}\label{LRnormalization}
Q^{L,[i]}&=&\sum_{\sigma_i}  M^{[i],\sigma_i \dag} M^{[i],\sigma_i}=1, \, \text{left-norm.} \\
Q^{R,[i]}&=&\sum_{\sigma_i}  M^{[i],\sigma_i} M^{[i],\sigma_i \dag}=1, \, \text{right-norm.}
\end{eqnarray}

The algorithm presented above not only allows the variational determination of ground states
but also the calculation of excited states. They can be constructed by finding
the lowest state in the (sub)space orthogonal to the space spanned by the
states already found~\cite{PhysRevB.73.014410}. In order to implement a corresponding update procedure for an MPS in this subspace we need to solve the generalized eigenvalue problem
\begin{equation}\label{projector}
P^{[i]}H_{\rm eff}^{[i]}P^{[i]\dagger} \nu^{[i]}=\epsilon^{[i]} P^{[i]}N_{\rm eff}^{[i]}P^{[i]\dagger}\nu^{[i]}
\end{equation}
where $P^{[i]}$ projects the effective Hamiltonian~(\ref{eq:eff-Hamiltonian-p}) and the effective normalization matrix~(\ref{eq:eff-norm-p}) into the relevant subspace.
I.e., the local MPS tensor must not only minimize the effective Hamiltonian, but must also
be constructed in such a way that the MPS $|\psi\rangle$ of a `new' excited state to be calculated is
orthogonal to all MPS $|\phi_k\rangle$ already calculated before: $\langle \psi|\phi_k\rangle=0$ for all $k$. Here, $k$ enumerates these states: $k=0$ for
the ground state, $k=1$ for the first excited state, \textit{etc}.

We denote the tensors of the states  $|\phi_k\rangle$
by $\Phi_k^{[i]}$ and the tensors of the `new' excited state by $M^{[i]}$. Then the local projection operator
$P^{[i]}$ must fulfill the condition
%begin{equation}\label{eq:proj}
$P^{[i]} y^{[i] \, \dagger}=0$
%end{equation}
with  $y$ a matrix of $k$ row vectors
\begin{eqnarray}\label{eq:proj1}
&&(y_k^{[i]})_{[\sigma_i a_i a_{i-1}]}=
%\sum_{a_N,a_N^{\prime}} \delta_{\sigma_i,\sigma_i^{\prime}}
\sum_{a_i^\prime a_{i-1}^\prime} (O_{R,k}^{[i]})_{(1, a_i, a_i^{\prime}),[1 a_N a_N^{\prime}]} \times \\
&&~~~~ (O_{L,k}^{[i]})_{[1 a_N a_N^{\prime}],(1, a_{i-1}, a_{i-1}^{\prime})} \cdot (\Phi_k^{[i]})^{\sigma_i}_{a_{i-1}^{\prime},a_i^{\prime}} \nonumber
%&&~~{\rm and}~~ y^{[i]}_k\cdot y^{[i]\dagger}_m=0~~{\rm if}~~k\neq m.\nonumber
\end{eqnarray}
obtained from the condition $\langle \psi|\phi_k\rangle=0$ for all $k$. The projection matrix
$P^{[i]}$ is determined by an orthogonalization of the matrix $y^{[i]}$.
The rank-6 `block' tensors $O_{L,k}^{[i]}$ and  $O_{R,k}^{[i]}$ are the contractions of all transfer tensors $E_1(M,\Phi_k)$  to the left or right of site $i$, respectively. (Again, summation over repeated indices is assumed
in Eq.~(\ref{eq:proj1}))

\section{SU(2)-symmetric MPS algorithm for PBC}\label{SU2mps}

We now introduce SU(2) symmetry into the formalism presented above: we assume an SU(2) symmetric Hamiltonian
and SU(2) symmetric states, i.e. the symmetry is not broken. This holds in 1D due to the Mermin-Wagner-Coleman theorem.
Technically, all tensor indices can then be chosen to be SU(2) invariants, i.e. they decompose into a total spin
(angular momentum) index $s$, a spin projection index $m_s$ as well as a degeneracy index $t$, which counts the number of times a specific SU(2) representation occurs, e.g. $\sigma=(s, t, m_s)$. For the physical indices of MPS we have $t=1$, however the spins in the bond indices are highly degenerate. In practice, we only need  moderate degeneracies $t<10$ for our calculations.

First we briefly introduce SU(2) symmetric MPS and discuss the construction of states with given total angular momentum  $J$. We eliminate all structural tensors from the algorithm in subsection~\ref{optalg} as well as several appendices.

\subsection{Construction of SU(2)-symmetric MPSs for PBC}

First we discuss how to construct SU(2) invariant states, i.e. states with total spin $J=0$. The set of MPS matrices defined in Eq.~(\ref{eq:MPS}) must be brought into an SU(2) symmetric form at every site $i$.

The structure of the rank-3 local tensor $M^{[i]}$ at position $i$ is determined by the Wigner-Eckart theorem~\cite{PhysRevB.86.195114} according to which the tensor decomposes into a degeneracy part $\mathcal{M}$ and a structural part $C$
\begin{equation}\label{mel}
M_{(jtm),(j^{\prime}t^{\prime}m^{\prime})}^{s m_s}=\mathcal{M}_{(j t),(j^\prime t^{\prime})} C_{m,m_s,m^\prime}^{j,s,j^\prime}
\end{equation}
with the structural part fixed by SU(2) symmetry and given in terms of the Wigner $3j$ symbol~\footnote{We use the conventions of Edmonds~\cite{edmonds} in this paper. For this reason we prefer to write the structural part in terms of $3j$ symbols instead of the closely related Clebsch-Gordan coefficients.}
\begin{equation}\label{c-factor}
C_{m_1,m_2,m_3}^{j_1,j_2,j_3}=(-1)^{j_1-m_1} \, \begin{pmatrix} j_1 & j_2 & j_3 \\ -m_1 & m_2 & m_3 \end{pmatrix}.
\end{equation}
The degeneracy part (also often called ``reduced tensor'') is denoted by a script letter while the corresponding full tensor is denoted by a roman letter. Here, for simplicity, we omit the position index $i$ (this index will be reintroduced whenever necessary). It is often convenient to use a combined index $\gamma=(j,t)$, i.e. $\mathcal{M}_{(jt),(j^\prime t^{\prime})}=\mathcal{M}_{\gamma,\gamma^{\prime}}$.

The degeneracy part $\mathcal{M}$ does not depend on the spin projections and contains the (variational) parameters of the state. Note that the reduced matrix elements of the MPS for which the spins $j,s,j^\prime$ do not fulfill the `triangle rule'
\[
\begin{cases}
|j-s| \leq j^\prime \leq j+s \\
j+s+j^{\prime} \,\,\, \text{is integer}
\end{cases}
\]
can be set to zero, as the corresponding $3j$ symbol vanishes under these conditions. This makes the reduced MPS matrices rather sparse, in fact, they have a banded block structure.

We will call the set of all $(jt)$ which characterize a reduced matrix its {\it degeneracy set} $d$, e.g. $d=\{(1/2,2),(3/2,4),(5/2,3)\}$. In practice, we want to choose the smallest possible set with few different spins $j$ and small degeneracies $t$. However, it must be noted that at present we have no method in order to choose this set algorithmically. Of course, for a large system we expect that the required degeneracies will be large. On the other hand the size of the degeneracy sets is strictly limited by the available computing resources. We will discuss this issue further in the following sections.

We now turn to the construction of covariant states with total spin $J \ne 0$.  To achieve this a fictitious non-interacting (local) spin $J>0$~\cite{McCulloch2001,Singh2010} is inserted into the system.
The fictitious spin is inserted at the site $N+1$, i.e. between site $N$ and site 1.  The tensor at this additional site takes the form
\begin{equation}
F_{(jtm),(j^{\prime}t^{\prime}m^{\prime})}^{J M}=\mathcal{F}_{(j t),(j^\prime t^{\prime})} C_{m, M, m^\prime}^{j, J, j^\prime},
\end{equation}
and the resulting MPS  has total spin 0. For completeness, we define $F^{0,0}=\mathds{1}$.
After optimization of the tensor network {\it with} the fictitious spin one needs to obtain the
covariant state $|J M\rangle$, which is given by
\begin{multline}\label{mpsrec}
|JM\rangle = \sum_{\{m_{s_i}\}} \textrm{Tr} (M^{[1],(s m_{s_1})}  \cdots M^{[N],(s m_{s_N})} F^{J,-M})\\ |m_{s_1} m_{s_2} \cdots m_{s_N} \rangle
\end{multline}
This state must be normalized if required. In practice, we seldom need to reconstruct
the state using Eq.~(\ref{mpsrec}), since the calculation of SU(2) invariant observables can be expressed in terms of the reduced tensors only as is shown further down in this section.

\subsection{The optimization algorithm for SU(2) symmetric MPS for PBC}\label{optalg}

In order to make best use of the symmetry, it is advantageous to introduce a reduced tensor structure not
only for the MPS but for \textit{all} other tensors of the algorithm as well. This enables to express the whole algorithm in terms of degeneracy tensors only, and spin projection indices and the structural parts of the MPS will
be completely eliminated from the algorithm. As a consequence, the computations are significantly more efficient,
and calculations with much larger virtual dimensions become feasible.

In the present section we will explain how the
PBC algorithm presented in section~\ref{standard} is rewritten in terms of reduced tensors only. Technical details will be
relegated to several Appendices. The results contained in these Appendices will enable a rather straightforward implementation of the
proposed algorithm.  For open boundary conditions a similar strategy was followed by McCulloch~\cite{McCulloch2007} but not described in much detail. However, for PBC
we face a number of differences, in particular, the `block' tensors, i.e., the products of transfer tensors, are rank-6 tensors for PBC, while they are rank-3 in the OBC implementation.

In the following, we will provide the general definition of reduced tensors
and exemplify the construction of reduced tensors for rank-4 $W$ tensors, which are the building blocks of MPO. We then go on to describe the variational algorithm for the determination of MPS in terms
of reduced tensors.

In order to introduce rank-$k$ SU(2) {\it invariant} tensors $T_{a_1,\ldots,a_k}$ each index has to be decomposed into a spin index, a degeneracy index,  and a spin projection index, e.g. $a_1=(j_1,t_1, m_{j_1})$ in the same way we did for the tensor indices of MPS tensors in Eq.~(\ref{mel}).
For $k \ge 3$ an element of a SU(2)-invariant rank-$k$ tensor may be obtained from the generalized Wigner-Eckart theorem~\cite{edmonds, PhysRevA.82.050301,PhysRevB.86.195114}
\begin{multline}\label{sutwotensor}
T_{a_1,a_2,\cdots,a_k}=\sum_{\substack{j_{e_1},t_{e_1},m_{e_1}\\\ldots\\j_{e_{k-3}},t_{e_{k-3}},m_{e_{k-3}}}} (\mathcal{T}^{j_{e_1},\cdots,j_{e_{k-3}}})_{\gamma_{a_1},\gamma_{a_2},\cdots,\gamma_{a_k}} \cdot \\ \cdot (Q_{j_{a_1},j_{a_2},\cdots,j_{a_k}}^{j_{e_1},\cdots,j_{e_{k-3}}})_{m_{a_1},m_{a_2},\cdots,m_{a_k},m_{e_1},\cdots,m_{e_{k-3}}}
\end{multline}
where $e_1,\ldots,e_{k-3}$ are intermediate indices~ to be summed over,~ $(\mathcal{T}^{j_{e_1},\cdots,j_{e_{k-3}}})_{\gamma_{a_1},\gamma_{a_2},\cdots,\gamma_{a_k}}$ is a reduced matrix element while $Q_{j_{a_1},j_{a_2},\cdots,j_{a_k}}^{j_{e_1},\cdots,j_{e_{k-3}}}$ is a rank-$k$ intertwiner of SU(2) (generalized Clebsch-Gordan coefficient).
Obviously, the tensor~(\ref{mel}) of an MPS is a special case of~(\ref{sutwotensor}) for $k=3$ (there are no intermediate indices, and the index $\sigma$ decomposes trivially as $\sigma=(s,1,m_s)$ or simply $\sigma=(s,m_s)$).

A rank-$k$ intertwiner may be decomposed into a product of $C$ factors defined in Eq.~(\ref{c-factor}). Different decompositions are possible and they may be represented as different coupling schemes as explained in more detail in Ref.~\cite{edmonds} and Ref.~\cite{PhysRevB.86.195114}. Here, we will choose coupling schemes which we find convenient for our purposes.
\begin{figure}
\unitlength1cm
\begin{picture}(6,3)(0,0)
 \put(2.,0)  {\includegraphics[width=1.5 cm]{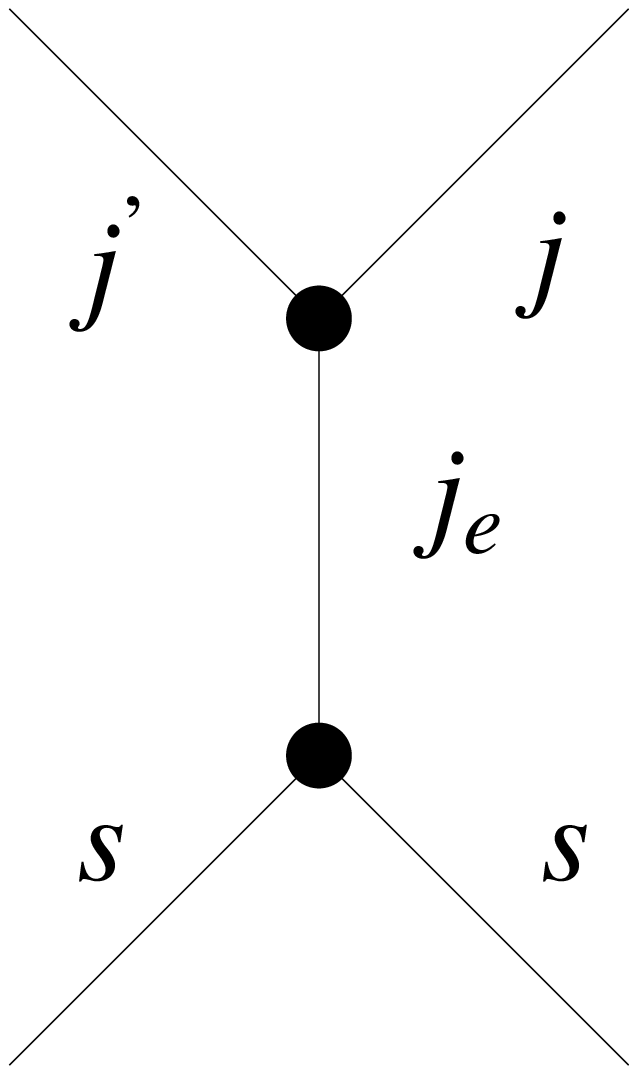}}
\end{picture}
\caption{\footnotesize Coupling scheme corresponding to the decomposition of a rank-4 $W$ tensor of an MPO Eq.~(\ref{reducedMPO}). Each vertex represents a $C$ factor or 3$j$ symbol. Summation over the indices of internal edges is implied.\label{fig:fstree}}
\end{figure}

Let us illustrate the construction of SU(2) invariant tensors using an MPO tensor $W$ as example.
Such tensors have rank-4, therefore we need one intermediate index $e=(j_e,m_e)$. The MPO has two physical indices $(s,m_s)$ and $(s,m_s^\prime)$ (assuming that at each site there are particles with the same spin $s$) and two virtual
indices $(j t m)$, $(j^\prime t^\prime m^\prime)$.
The $W$ tensor is then  decomposed as
\begin{multline}\label{reducedMPO}
W_{{(j t m),(j^\prime t^\prime m^\prime)}}^{(s m_s),(s m_s^\prime)}=\\
 =\sqrt{2s+1} \sum_{j_e,m_e} \mathcal{W}_{(j t),(j^\prime t^\prime)}^{j_e} \, C_{m,m_e,m^{\prime}}^{j,j_e,j^{\prime}} \, C_{m_s,m_e,m_s^{\prime}}^{s,j_e,s}
\end{multline}
with a rank-3 reduced tensor $\mathcal{W}_{\gamma,\gamma^{\prime}}^{j_e}$
(the factor $\sqrt{2s+1}$ is introduced for convenience in order to free the reduced unity tensor from an $s$ dependence).
Our definition of the reduced tensor corresponds to the coupling scheme shown in Fig.~\ref{fig:fstree}. We note that $t_e=1$ because fusion of two equal spins $s$ gives a non-degenerate spin. Due to the restrictions imposed by the structural $C$ factors as well as the high sparseness of the MPO in its full form, many reduced tensor elements can be chosen to be zero, and the reduced tensor $\mathcal{W}$ assumes a characteristic sparse structure.
This is demonstrated by the reduced tensor representations of the Hamiltonians needed in the present paper given in Appendix~\ref{heisred}.
We also provide a reduced representation for the $H^2$ operator in Appendix~\ref{Hsquare}.

As was already stated, the $W$ tensors of many important operators can be obtained {\it exactly} as well as their reduced tensor representation $\mathcal{W}$. This representation is characterized by a degeneracy set $d=\{(j_0,t_0),\ldots (j_n,t_n)\}$, e.g. for the BBH Hamiltonian one finds $d=\{(0,3),(1,1),(2,1)\}$ as is shown in Appendix~\ref{heisred}.

Similarly, one can expand higher rank tensors like the block tensors $H_R$ and $H_L$ into reduced tensors.
Details are given in Appendix~\ref{reducedTensors}.
As a consequence, the optimization (update) step~Eq.~(\ref{eq:eigenproblem-p}) can be formulated as a generalized eigenvalue problem
in terms of the reduced effective Hamiltonian $\mathcal{H}_{\rm eff}$ and the reduced effective normalization matrix  $\mathcal{N}_{\rm eff}$ (both given explicitly in Eq.~(\ref{Oeff}))
\begin{equation}\label{redupdate}
\mathcal{H}_{\rm eff} \, \mathcal{\nu}=\epsilon \, \mathcal{N}_{\rm eff} \, \mathcal{\nu}.
\end{equation}
We stress again that the reduced normalization matrix arises due to PBC and both  $\mathcal{H}_{\rm eff}$ and $\mathcal{N}_{\rm eff}$ change at each iteration step.
The updated matrix $\mathcal{M}$ is then obtained from a suitable partition of the vector $\nu$ into a matrix, $\mathcal{M}_{\gamma,\gamma^{\prime}}=\mathcal{\nu}_{[\gamma^{\prime}\gamma]}$ as well as zeros for those  tensor elements which do not fulfill the triangle rule.

Details of the derivation of the optimization step in terms of reduced tensors are given in Appendix~\ref{redeig}. A reduced form of the projection operator $P$ defined
in Eq.~(\ref{projector}) is also given there. This enables the calculations of excited states in different spin sectors. The regauging step (Eq.~(\ref{LRnormalization})) in terms of reduced tensors is briefly described in Appendix~\ref{regauging}. A first demonstration that the proposed algorithm works as expected is given in Appendix~
\ref{spin-1/2}.

\subsection{Calculation of observables}

If the operator $O$ of an observable is SU(2) invariant, its expectation value Eq.~(\ref{eq:matrixe}) can be calculated using reduced tensors only. In the present paper we do not consider covariant operators.
Using Eq.~(\ref{eq-MPO}) for the operator and Eq.~(\ref{mpsrec}) for the wavefunction $|JM\rangle$, one can write
\begin{multline}
\langle J^\prime M^\prime|O|JM \rangle=\delta_{J^\prime J} \, \delta_{M^\prime M}\times \nonumber\\
{\rm Tr} [(\sum_{\sigma_1,\sigma_1^{\prime}} W^{[1],\sigma_1,\sigma_1^\prime} \otimes M^{[1],\sigma_1 *} \otimes M^{[1],\sigma_1^{\prime}}) \times \nonumber\\
\times \cdots \times (\sum_{\sigma_N,\sigma_N^{\prime}} W^{[N],\sigma_N,\sigma_N^\prime} \otimes M^{[N],\sigma_N *} \otimes M^{[N],\sigma_N^{\prime}}) \times \nonumber \\
\times (\mathds{1} \otimes F^{J,-M*} \otimes F^{J,-M})]. \nonumber
\end{multline}
The expectation values of the SU(2) invariant operator are equal for different spin projections: $\langle JM|O|JM\rangle=\langle JM^{\prime}|O|JM^{\prime}\rangle$. Therefore, we can write
\begin{eqnarray}
\langle J|O|J \rangle&=&\frac{1}{2J+1} \sum_M \, \langle JM|O|JM\rangle=\nonumber\\
&=&{\rm Tr} (E_W^{[1]}(M,M) \cdots E_W^{[N]}(M,M) E_1(F,F))\nonumber\\
&=&\sum_{j_{e_1},j_{e_3}} \frac{1}{2j_{e_1}+1} \, \textrm{Tr} \, {\mathcal{B}_L}^{j_{e_1},j_{e_1},j_{e_3}},
\end{eqnarray}
if the MPS tensor $F$ at the fictitious site is normalized such that $Q^R=\mathds{1}$. For simplicity we did not label
the MPS matrices with their corresponding local spin index. In the last line we have represented the expectation value
as a reduced left block tensor as defined in Appendix~\ref{reducedTensors}. It is this represenation which we use
for a recursive calculation of expectation values using the formulas provided in Appendix~\ref{reducedTensors}.

An example of an SU(2) invariant operator is the dimerization operator~(\ref{dimerorder}). On the other hand, the string correlator~(\ref{stringorder}) is not SU(2) symmetric due to the parity factor $\prod_{j=i+1}^{i+l-1} \, \exp(i\pi s_j^z)$.  For such  operators a full MPS needs to be reconstructed using~(\ref{mpsrec}), and the MPO in its full form must be used.

\section{Bilinear-biquadratic spin-1 Heisenberg model with PBCs}\label{applications}

In this section we present a numerical study of the spin-1 bilinear-biquadratic Heisenberg (BBH) model Eq.~(\ref{eq-bilbiq}). This model has been studied extensively in recent years, and we have given a number of pointers to the literature in the Introduction. The present study is distinguished by its particular treatment of SU(2) symmetry in periodic systems. With moderate computational effort and relatively small numbers of variational parameters we obtain results of high precision.
We present results for the low lying spectrum of the BBH model as well as for selected correlation functions.
In order to do calculations using the proposed algorithm one needs to choose an appropriate degeneracy set for
the virtual spins. The larger this set the larger is the computational effort. An algorithmic  procedure for automatic selection of a suitable degeneracy set is presently under development.

\begin{figure}
\unitlength 1cm
\includegraphics[width=0.4\textwidth]{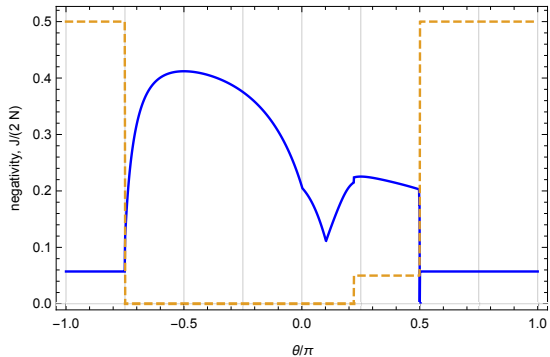}
\caption{\footnotesize (color online) The negativity (full blue) of the spin-1 bilinear-biquadratic Heisenberg ring of $N=10$ spins as a function of $\theta$ obtained using exact diagonalization. The total spin $J$ of the ground state (dashed orange) is $J=0$ for $-0.75<\theta/\pi \lesssim 0.22$, $J=1$ for $0.22 \lesssim \theta/\pi <0.5$ and $J=N$ for $0.5 < \theta/\pi <1.25$.  We plot the quantity $J/(2N)$.
\label{negat10}}
\end{figure}

In order to select the degeneracy sets we use the entanglement structure for guidance. Obviously,
for highly entangled states one needs large degeneracy sets with many virtual spins.
In order to estimate the entanglement of the states at different $\theta$, we use exact diagonalization for a system of $N=10$ spins and determined the negativity~\cite{PhysRevA.65.032314} of the ground state to quantify nearest-neighbor  qutrit-qutrit entanglement (see Fig.~\ref{negat10}). The negativity $N$ is obtained from the reduced 2-qutrit density matrix $\rho$,
$$
N(\rho)= \frac{1}{2}(\|\rho^{T_1}\| -1),
$$
where $T_1$ denotes the partial transpose of $\rho$ with respect to the first qutrit and $\|.\|$ the trace norm.

It is obvious from Fig.~\ref{negat10} that bipartite entanglement is largest in the dimerized phase, while in the Haldane phase and the critical phase the states are less strongly bipartite entangled. Therefore, we cautiously expect that the largest computational resources will be needed in the dimerized phase. At the AKLT point $\theta=\arctan\frac{1}{3}$ there is a local entanglement minimum. At this point we just need the trivial virtual spin representation $\{(1/2,1)\}$ for a quasi-exact calculation for any system size. The bi-partite entanglement vanishes at the point $\theta=\pi/2$ (which happens even for large systems).
%Here, the ground state is a product state.
At $\theta=-3\pi/4$ the negativity vanishes as well, however, at this point the ground state is in fact highly entangled. It resembles a state similar to the Greenberger-Horne-Zeilinger (GHZ) state, which is maximally entangled but measures zero nearest-neighbor entanglement.
%Meanwhile, in the ferromagnetic phase the ground state has spin $J=N$, and the trivial degeneracy set $\{(N/2,1)\}$ is needed for calculations.

Our choice for the virtual spin representations and their degeneracies is partly guided by these entanglement results. The negativity shows characteristic singularities at various points, foreshadowing clearly already for $N=10$ the phase structure one observes in the thermodynamic limit. Interestingly, there is a weak singularity at the Heisenberg point $\theta=0$, where there is no phase transition in the thermodynamic limit.

Together with the bipartite entanglement we also plot in Fig.~\ref{negat10} the quantity $J/(2 N)$ ($J$ is the total spin of the ground state) for $N=10$. In the parameter region of the critical phase $J=1$, while it is zero in all other phases except the ferromagnetic phase.

The MPSs we construct are eigenstates of the BBH Hamiltonian as well as eigenstates of the total angular momentum operator (therefore we label the states with angular momentum $J$). However, they are not necessarily eigenstates of quasi-momentum operator as well. Nevertheless, if the constructed eigenstates are non-degenerate (apart from the SU(2) degeneracies), we determine in some cases the quasi-momentum of the constructed states and label the states and their energies with the momentum label $p$.
%As already pointed out, the determined states either have well-defined momentum or they are superpositions of degenerate eigenstates with different momenta.
The operator $T_n$ of the translation over $n$ sites acts on an eigenstate with well-defined momentum $p$ as
$$
T_n |\psi\rangle=e^{-ipn}|\psi\rangle.
$$
Therefore, for such states $p=2\pi n_p/N$ ($n_p=0,\ldots,N-1$) can be determined from the expectation value of the translation operator, $\langle \psi|T_n |\psi\rangle=e^{-ipn}$, which is easily calculated in the MPS framework
using Eq.~(\ref{eq:matrixe}) and a cyclic shift of the MPS tensors in $|\psi\rangle$.

\subsection{Ground state energy}

\begin{figure}%[t!]
\unitlength 1cm
\includegraphics[width=0.4\textwidth]{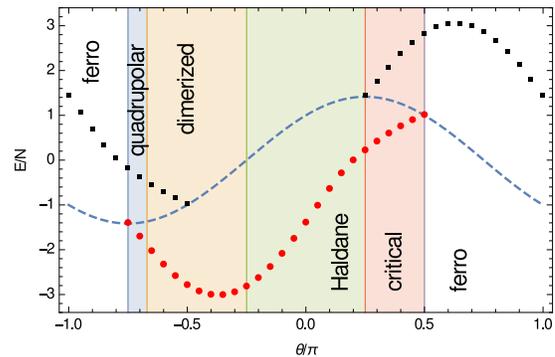}
\caption{\footnotesize (color online) Lowest (red dots) and highest (black dots) energy per site with $J=0$ as a function of $\theta$  for the spin-1 bilinear-biquadratic Heisenberg ring with $N=100$ spins. The lowest energy in the $J=N$ sector (blue dashed line) $E/N=\cos\theta+\sin\theta$ is independent of the system size. In the ferromagnetic phase this energy corresponds to the `global' ground state, while in the region $-\frac{1}{2}<\theta/\pi <\frac{1}{4}$ this energy corresponds to the `global' highest excited state.
\label{energy}}
\end{figure}

Fig.~\ref{energy} shows results for the ground state energies per site in the $J=0$ and $J=N$ sectors for $N=100$ spins as a function of the parameter $\theta$. The `global' ground state,  i.e. the lowest state of all spin sectors, is $J=0$ except in  the {\it critical phase} ($\frac{1}{4} <\theta/\pi<\frac{1}{2}$), where the `global' ground state is $J=1$ for $N=100$, and the {\it ferromagnetic region}. Due to the symmetry $H(\theta)=-H(\theta+\pi)$ the minimal and maximal eigen-energies are related by $E_{\rm min}(\theta)=-E_{\rm max}(\theta+\pi)$ (this maximal energy is also shown in Fig.~\ref{energy}). With few exceptions we used the representation $\{(1/2, 6), (3/2, 6), (5/2, 5)\}$ in order to produce this plot.
Note that results for $N=100$ are rather close to the thermodynamic limit.

At $\theta/\pi=-\frac{3}{4},-\frac{1}{2},-\frac{1}{4},\frac{1}{4},\frac{1}{2}$ exact ground state energies per site are known from Bethe Ansatz calculations. We compare with these results in Table~\ref{dataN100CP} and find rather good agreement.
It is interesting that our numerical results are obtained using half-integer virtual spins only (we obtain similar values with integer virtual spins, however, at a somewhat larger numerical cost).
Precise calculations are most resource intensive for the two SU(3) symmetric points, $\theta=\pi/4$ and $\theta=-3\pi/4$.  In particular, virtual spins $j=7/2$ are necessary in the degeneracy set $d$ in the vicinity of  $\theta=-3\pi/4$.

\begin{table}
\begin{tabular}{|c|c|c|c|c|}
  \hline
  $\theta/\pi$ & $E/N$ & $E_T/N$ & $\Delta E/E_T$ & $d$ \\
  \hline
  0.25  & 0.21058604015 &  & $ $ & $\{6,6,5,0\}$ \\
  \hline
  0.5   & 1.00000004283 & 1.0           & $4.3 \cdot 10^{-8}$ & $\{1,1,1,0\}$ \\
  \hline
  -0.75 & -1.4100386467 & -1.4142135624 & $3.0 \cdot 10^{-3}$ & $\{3,3,3,2\}$ \\
  \hline
  -0.5  & -2.7967884870 & -2.7969307339 & $5.1 \cdot 10^{-5}$ & $\{6,6,5,0\}$ \\
  \hline
  -0.25 & -2.8286690852 &  & $ $ & $\{7,7,6,0\}$ \\
  \hline
\end{tabular}
  \caption{\footnotesize Ground state energy per site for the bilinear-biquadratic spin-1 Heisenberg ring of $N=100$ sites at various values of the parameter $\theta$. $E_T$ are finite-size predictions for $\theta/\pi=0.5,-0.75,-0.5$ (e.g.,~\cite{PhysRevB.42.754}). Infinite-size Bethe ansatz results for $\theta/\pi=0.25$~\cite{PhysRevB.12.3795,Nomura1991} $E/N=0.20986075311$ and $\theta/\pi=-0.25$ $E/N=-2.8284271247$~\cite{Takhtajan1982}. The spin representations  $d=\{(1/2,x_1),(3/2,x_2),(5/2,x_3),(7/2,x_4)\}$ are simply denoted as $\{x_1,x_2,x_3,x_4\}$.
  \label{dataN100CP}}
\end{table}

In the {\it ferromagnetic region} $\frac{1}{2} < \theta/\pi < \frac{5}{4}$
the bilinear-biquadratic spin-1 Heisenberg model has a $2N+1$ degenerate spin-$N$ ground state for any system size $N$. It is possible to work out the ground state energy analytically within the proposed MPS algorithm as the virtual spin configuration is simply $\{(N/2,1)\}$. We demonstrate this in Appendix~\ref{marginalspin}.

\subsection{Low lying spectrum}

While it is possible to extract much information on the phase structure from an analysis of the ground state properties, in the present paper we will concentrate on the low lying excitation spectrum, and we will discuss the spectrum at selected points in the phase diagram. We will emphasize characteristic differences of the low lying spectrum in different phases which show up already
for finite systems.

\begin{figure}%[t!]
\unitlength 1cm
\includegraphics[width=0.4\textwidth]{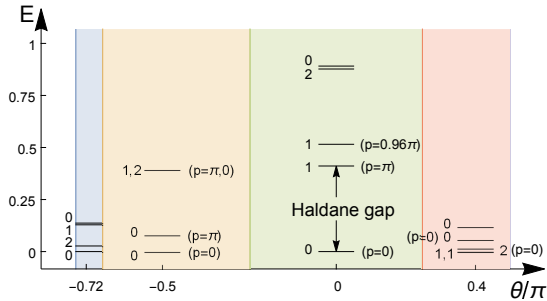}
\caption{\footnotesize (color online) Low lying spectrum of the bilinear-biquadratic Heisenberg ring of $N=50$ spins in the parameter range $-3/4 < \theta/\pi < 1/2$. Two lowest spin-0 states, one or two spin-1 states and one spin-2 state are shown. The ground state energies are shifted to $E=0$. The ticks mark the values of $\theta$ where the spectrum is calculated. The color code for the various phases is introduced in Fig.~\ref{phased}.
\label{spectrum}}
\end{figure}

The spectrum for a system of 50 sites within the range $-\frac{3}{4} < \theta < \frac{1}{2}$ is presented in Fig.~\ref{spectrum}. We include the lowest two spin-0 states, one or two spin-1 states and one spin-2 state.  The states are labelled by their momentum $p$ in some cases.
We discuss the spectra in the various phases shown in Fig.~\ref{spectrum} from right to left:

Characteristically, in the {\it critical phase} ($\frac{1}{4} < \theta/\pi < \frac{1}{2}$) the ground state for 50 spins is doubly degenerate spin-1.
There is small gap to the lowest spin-2 state, and there are slightly larger gaps to two lowest spin-0 states.
Consequently, the lowest excitation is expected to be quadrupolar in this phase, which confirms predictions for spin chains in~\cite{PhysRevB.74.144426}. Our numerical results for the  ground state for the systems with 50 or 100 sites at $\theta=0.4\pi$ agree quantitatively very well with predictions in Ref.~\cite{PhysRevB.44.11836} obtained by extrapolation of the results for relatively small systems.

The ground state in the entire {\it Haldane phase} ($-\frac{1}{4} < \theta/\pi < \frac{1}{4}$) is $J=0$, and the lowest excited state is $J=1$. The gap between these two states remains finite in the thermodynamic limit (Haldane gap).
Numerical data for the $J=0$ state and $J=1$ state for a purely bi-linear Heisenberg ring with $N=100$ sites are given in Table~\ref{dataN100}. From these results one obtains the Haldane gap $\Delta=0.41096$ as compared to $\Delta=0.41048$ given in Ref.~\cite{PhysRevB.85.035130, PhysRevB.85.100408}. At the given precision $N=100$ cannot be distinguished from infinite systems. The calculated energy of the $J=0$ state agrees well with the results from previous studies~\cite{PhysRevB.72.180403,PhysRevB.81.081103}.
\begin{table}
\begin{tabular}{|c|c|c|c|c|}
  \hline
   $J$ & $E/N$ & $E_{\rm C}/N$ & $\Delta E/E_{\rm C}$ &  $d$ \\
  \hline
  0 & -1.4014840187 & -1.4014840390~\cite{PhysRevB.72.180403,PhysRevB.81.081103} &  1.4 $\cdot 10^{-8}$ & $\{6,6,6\}$ \\
  \hline
  1 & -1.3973744069 & -1.3973792452~\cite{PhysRevB.85.035130,PhysRevB.81.081103} &  $3.5 \cdot 10^{-6}$ & $\{9,9,9\}$ \\
  \hline
\end{tabular}
  \caption{\footnotesize Lowest energies per site of the bi-linear spin-1 Heisenberg ring of $N=100$ sites. The spin representations used are $d=\{(1/2,x_1),(3/2,x_2),(5/2,x_3)\}$, which we denote simply as $\{x_1,x_2,x_3\}$. The Haldane gap is $\Delta E=0.41096$ (best estimate: $\Delta E=0.41048$). $E_{\rm C}$ are results from other numerical calculations as indicated by the citations. $E_C$ for $J=1$ is deduced from best estimates for the Haldane gap and for the energy of the $J=0$ state.
  \label{dataN100}}
\end{table}
The data presented in Fig.~\ref{spectrum} refer to 50 sites. It is observed that at $\theta=0$ the first excited state in the spin-0 sector lies much
above the lowest spin-1 state, and it is nearly degenerate with the lowest spin-2 state (see also \cite{PhysRevB.85.100408}).
%Our numerical results
%%for the lowest spin-2 state and the first excited state in the spin-0 sector
%for the Heisenberg point $\theta=0$ agree well with the results obtained for 100 sites in~\cite{PhysRevB.85.035130} and for infinite-size calculations presented in~\cite{PhysRevB.85.100408}.

The {\it dimerized phase} ($-\frac{3}{4}< \theta/\pi<-\frac{1}{4}$) has been the subject of many investigations in the history of the BBH model, and we will address the calculation of the dimerization in the next subsection. This phase is characterized by the fact that the first excited spin-0 state is lower than the lowest spin-1 and spin-2 states. The spin-1 state is lower than the spin-2 state for $\theta/\pi>-\frac{1}{2}$ but above this state  for $\theta/\pi<-\frac{1}{2}$.  At $\theta=-\pi/2$ the lowest spin-1 and spin-2 states form a degenerate pair (see also ~\cite{PhysRevB.47.872}). In the thermodynamic limit, the two lowest $J=0$ states are degenerate, however with a finite gap to the lowest $J=1$ or $J=2$ states. As shown in Table~\ref{dataBQ} the calculated energies of the three lowest states at the biquadratic point $\theta=-\pi/2$ agree very well with the Bethe Ansatz calculations of~S{\o}rensen and Young~\cite{PhysRevB.42.754}.
According to the heuristic entanglement analysis presented
in the introduction to this section we expect calculations to be rather ``hard'' at this point. And indeed, we need large degeneracy sets in order to achieve reasonable precision.
\begin{table}[t]
\begin{tabular}{|c|c|c|c|c|}
  \hline
  $J$ & $E/N$       & $E_{\rm BA}/N$ & $\Delta E/E_{\rm BA}$  & $d$ \\
  \hline
  0 & -2.7974653891 & -2.7974930571  & 1.0 $\cdot 10^{-5}$    & $\{8,8,7,0\}$ \\
  \hline
  0 & -2.7958843262 & -2.7959185488  & 1.2 $\cdot 10^{-5}$    & $\{8,8,7,0\}$ \\
  \hline
  1 & -2.7897606968 & -2.7899887203  & 8.2 $\cdot 10^{-5}$    & $\{8,8,8,0\}$ \\
  \hline
  2 & -2.7897789933 &                &                        & $\{6,6,6,6\}$ \\
  \hline
\end{tabular}
  \caption{\footnotesize The energies per site of four low-lying states of the purely biquadratic ($\theta=-\pi/2$) spin-1 Heisenberg ring of $N=50$ sites. The Bethe ansatz results $E_{\rm BA}$ are taken from~\cite{PhysRevB.42.754}. The spin representations used are $d=\{(1/2,x_1),(3/2,x_2),(5/2,x_3),(7/2,x_4)\}$, which we denote simply as $\{x_1,x_2,x_3,x_4\}$. The calculated gap between the two $J=0$ states is $\Delta_{00}=0.0791$ (Bethe Ansatz: $\Delta_{00}= 0.0787$). The calculated gap between the spin-0 ground state and the spin-1 state is $\Delta_{01}= 0.3852$
  (Bethe Ansatz: $\Delta_{01}= 0.3752$).
  \label{dataBQ}}
\end{table}

In the parameter region of the debated {\it nematic phase} close to $\theta=-3\pi/4$  PBC rings of $N \leq 16$ sites were investigated in~\cite{PhysRevB.51.3620}. We extend this work for systems of up to 50 sites. We observe
that the lowest spin-2 state is significantly lower than the first excited spin-0 state as well as the lowest spin-1 state.
%Principally the same picture is observed also for a system of 10 sites.
As a consequence, quadrupolar correlations increase strongly in this region as pointed out in~\cite{PhysRevB.74.144426}.

One may be tempted to conjecture the existence of the separate phase in this region from our results. However, in order to do so one must study the behaviour of the spectrum  in the thermodynamic limit, which we do using  finite-size scaling analysis for the gaps $\Delta_{00}$ and $\Delta_{02}$.
The conjecture of Ref.~\cite{PhysRevB.43.3337} about the existence of the gapped nematic phase (i.e., $\Delta_{00}^{\infty}>0$ and $\Delta_{02}^{\infty}=0$) was rejected in earlier works~\cite{PhysRevB.51.3620,PhysRevLett.95.240404,PhysRevLett.98.247202}. In fact, these calculations  suggest that $\Delta_{00}^{\infty}=0$. However, there exists still the possibility that $\Delta_{02}^{\infty}=0$ in a narrow region of $-\frac{3}{4}<\theta/\pi \lesssim -0.7$. Earlier DMRG calculations for OBC~\cite{PhysRevB.74.144426}  were not able to resolve extremely small values of $\Delta_{02}^{\infty}$.

From a finite size scaling analysis we obtain $\Delta_{00}^{\infty} \lesssim 2 \cdot 10^{-4}$ at $\theta/\pi=-0.72$. This suggests that the gap between the two lowest spin-0 states closes in agreement with earlier calculations. In fact,  $\Delta_{00}(N)$ can be fitted very well by a power law $\Delta_{00}(N)=BN^{-\alpha}$. In line with this, the dimerization remains finite, $D_{\infty}=(4 \pm 2) \cdot 10^{-3}$,  as discussed in more detail in the following subsection. In contrast,  $\Delta_{02}(N)$ shows exponential behavior. From a fit to our data we obtain a gap between the lowest spin-0 and the spin-2 states of size $\Delta_{02}^{\infty} \simeq 0.009$ at $\theta=-0.72\pi$. Moreover, we observe a monotonic decrease of the scaled gap $N\Delta_{00}(N)$ and a monotonic increase of the scaled gap $N\Delta_{02}(N)$ with increasing of $N$, which also indicates the absence of a nematic state at this $\theta$. However, as pointed out in~\cite{PhysRevB.72.054433}, the possibility of the existence of such a state even closer to $\theta/\pi=-\frac{3}{4}$ cannot be excluded. Calculations closer to the ferromagnetic region for larger systems are presently under way.

\subsection{Correlation functions}

\begin{figure}%[t!]
\unitlength 1cm
\includegraphics[width=0.4\textwidth]{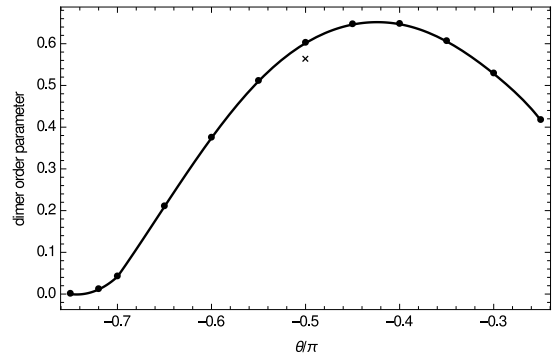}
%\hskip1cm
%\includegraphics[width=0.4\textwidth]{string100.eps}
\caption{\footnotesize (color online)  Dimer order parameter Eq.~(\ref{dimerorder}) calculated for the bilinear-biquadratic Heisenberg ring of $N=50$ sites as a function of $\theta$. $D(N) \simeq 0.6015$ at $\theta=-\pi/2$. The theoretical prediction for the infinite system at $\theta=-\pi/2$ is indicated by a cross~\cite{Xian1993,Baxter1973}, and our extrapolated value $D_{\infty} \simeq 0.568$ is in good agreement with theory.
\label{dimer}}
\end{figure}

Finally, we also calculate a few physically interesting correlations in the BBH model. In the area $-3/4 < \theta/\pi < -1/4$ we determine the dimerization correlator
\begin{equation}\label{dimerorder}
D=\frac{1}{N} \, \sum_i (-1)^i [\cos \theta \, \vec s_i \otimes \vec s_{i+1}+\sin \theta \, (\vec s_i \otimes \vec s_{i+1})^2],
\end{equation}
shown for a system of $N=50$ sites in Fig.~\ref{dimer}.
In fact, due to translational symmetry, the ground state as well as the excited states of a finite-size PBC ring are {\it not} dimerized. In order to calculate the dimerization we take a symmetric/antisymmetric superposition of the lowest two $J=0$ states with different momenta. These two states are separated by a very small gap for finite systems, and they should develop into the degenerate doublet in the thermodynamic limit.
The dimer order parameter calculated in this way is
$$
D=\frac{1}{2}\langle 0^{(0)} \pm 0^{(\pi)}|D|0^{(0)} \pm 0^{(\pi)} \rangle=\pm \langle 0^{(0)}|D|0^{(\pi)} \rangle.
$$
The dimerization $D$ can be calculated analytically for $\theta=-\pi/2$ in the thermodynamic limit, where the BBH model can be mapped to a spin-1/2 XXZ model. According to Refs.~\cite{Xian1993,Baxter1973} $D=\frac{\sqrt{5}}{2} \prod_{n=1}^\infty \tanh^2(n \, {\rm arccosh} \frac{3}{2}) \simeq 0.5622$, which agrees with the Monte-Carlo result of Ref.~\cite{PhysRevB.42.754}.

We fitted our results for 30, 40 and 50 sites at $\theta=-\pi/2$ by a function given in Eq.~(6) of~\cite{PhysRevB.72.054433}:
$$
D(N)=D_{\infty}+c\,N^{-1}\,\exp(-N/2\xi).
$$
The fit gives $D_{\infty} \simeq 0.568$ in good agreement with the theory, and $\xi \simeq 20.2$. Calculations for larger systems (about 100 sites) are needed to obtain a more precise result.

The dimer order parameter~(\ref{dimerorder}) was calculated in~\cite{PhysRevB.51.3620} for finite chains of up to 48 sites. In such OBC calculations there is no translational symmetry, and the degenerate doublet is mixed automatically. For systems of up to 50 sites the results for PBC are always slightly larger than for OBC.
There are also calculations of the dimerization for spin rings in Ref.~\cite{PhysRevB.73.014410},
%Unfortunately, this reference is not very clear as to how the dimerization is calculated. In particular, the authors do not seem to have appreciated that the dimerization in the ground state of a finite ring is zero due to translational invariance for any system size.
however for the four-point correlator $\langle 0^{(0)}|D^2|0^{(0)} \rangle$. Our extrapolated result for $\theta=-0.65\pi$ is consistent with the result of~\cite{PhysRevB.73.014410}.

\begin{figure}%[t!]
\unitlength 1cm
\includegraphics[width=0.4\textwidth]{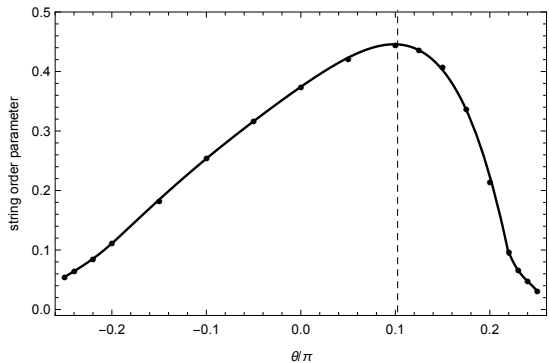}
\caption{\footnotesize (color online) String order parameter calculated for the bilinear-biquadratic Heisenberg ring as a function of $\theta$ (the dashed line denotes the AKLT point). The system size is $N=100$ and the string length is $l=30$ (to secure sufficient length of the string as well as the rest of the system). The value for $\theta=0$ is $g(l)\simeq 0.374330$ in a very good agreement with infinite-size calculations~\cite{PhysRevB.48.3844}.
\label{string}}
\end{figure}

Finally we come back again to the Haldane phase:
This phase is characterized by the presence of a nonzero string correlator of the ground spin-0 state~\cite{1402-4896-1989-T27-027},
\begin{equation}\label{stringorder}
g(l)=-\langle s_i^z (\prod_{j=i+1}^{i+l-1} \, \exp(i \pi s_j^z)) s_{i+l}^z\rangle,
\end{equation}
that does not decay in the limit $N,l \rightarrow \infty$.
Fig.~\ref{string} shows the string correlator $g(l)$ as a function of $\theta$.
%(we extend the investigation of~\cite{PhysRevB.44.11789,PhysRevB.46.13914,Hatsugai1992} to the entire Haldane phase and to a much larger system).
The system size is $N=100$, and the string length is taken as $l=30$. This choice allows sufficient length of both the string and the rest of the system, in order to resemble the infinite-system properties (numerical results indicate that $g(l)$ is constant to a high degree for $20 \le l \le 40$). Our results in the range $-0.2 \le \theta/\pi \le 0.2$ agree quantitatively very well with infinite-size calculations presented in~\cite{PhysRevB.53.3304}.

Our numerical calculations indicate that long-ranged string order is present in the Haldane phase only (however, it is numerically hard to obtain precise values at the boundaries of the Haldane phase).
The string correlator for the AKLT point ($\theta=\arctan{1/3}$) can be calculated exactly~\cite{PhysRevLett.59.799}: $g(\infty)=4/9$, which is reproduced extremely well in our calculation for $N=100$ sites. The curve has the maximum at this point.
The best estimate for a string correlator for the bilinear Heisenberg spin-1 model ($\theta=0$) is $g(\infty) \simeq 0.374325$~\cite{PhysRevB.48.3844}. Our result $g(30)=0.374330$ calculated for a system size of $N=100$ spins agrees very well with this value.

\section{Conclusions}\label{conclusions}

In the present work we developed an algorithm for SU(2) symmetric matrix product states (MPS) with periodic boundary conditions (PBC) that involves only reduced tensor elements. It was applied to a study of the lowest-lying states of the spectrum of the spin-1 bilinear-biquadratic Heisenberg model of up to 100 sites. The characteristic differences of the spectrum in the various phases of the model were stressed. Dimerization and string order were studied in the dimerized and Haldane phase, respectively.
Our results  agree rather well with previous studies based on DMRG or Bethe Ansatz calculations, and  we could extend previous studies to a more complete coverage of the full parameter range, in particular in the dimerized phase close to the transition to the ferromagnetic phase. We confirm the absence of a nematic phase for $\theta/\pi >-0.72$ within our numerical precision.

The precision of the results we can achieve with our implementation of the algorithm depends on the degeneracy set $d$ we choose for the representation of the virtual spin indices of the MPS. Due to fact that we have eliminated all
spin projection indices, our algorithm achieves rather large `effective' virtual spin dimensions in comparison to a
non-symmetric implementation. E.g. a state with reduced MPS dimension $m=30$ described by the degeneracy set $d=\{(0,6),(1/2,6),(1,6),(3/2,6),(2,6)\}$ corresponds to an effective MPS
virtual dimension $m=90$, which is already rather large for a PBC calculation. Moreover, due to the sparse structure
of each MPS reduced tensor, such a state is described by only 288 complex parameters per site compared to 16200 complex parameters
per site necessary to specify a non-symmetric state. The number of parameters is also roughly one order smaller than for a U(1) symmetric state. For this state about 98\% of the parameters used in a non-symmetric calculation are actually zero due to symmetry. As a consequence, the numerical effort for the solution of the generalized eigenvalue problem at each update step is reduced correspondingly. In fact, we suspect that the states
determined without explicit consideration of the symmetry are incorrect even though their energy is determined precisely.
This needs to be investigated in more detail in future work.

Of course, the numerical effort to be expected grows with the size of the chosen degeneracy set $d$. And at this stage
this set is fixed from the start of our calculation. It would be desirable that the algorithm dynamically chooses
this set according to suitable algorithmic criteria. The implementation of such a procedure is under way using
ideas presented in Ref.~\cite{PhysRevB.72.180403} and discussed for PBC in Ref.~\cite{PhysRevB.93.054417}.

\begin{acknowledgments}
We thank Briiissuurs Braiorr-Orrs for discussions.
Mykhailo V. Rakov thanks Physikalisch-Technische Bundesanstalt for financial support during short visits to Braunschweig.
\end{acknowledgments}

\appendix

\section{Reduced representation of MPOs}\label{heisred}

In Eq.~(\ref{reducedMPO}) we have defined the $W$ tensors of MPO in terms of reduced tensors $\mathcal{W}$. Here, we will present explicit results for the reduced tensors needed in the present study. In fact, all formulas in this Appendix are valid for arbitrary spin-$s$.

In order to construct the reduced tensor $\mathcal{W}_H$ for the spin-$s$ BBH Hamiltonian, we rewrite this Hamiltonian in terms of the tensors $C_{m_s, m_e, m_s^{\prime}}^{s,j_e,s}$, i.e. replace the spin matrices by $C$ tensors. In general, we would expect terms with $j_e=0,1,\cdots,2s-1,2s$, however one only finds non-zero terms with $j_e\leq 2$ for this Hamiltonian.
The corresponding $W_H$ tensors have dimension $11\times 11\times s \times s$, which can be grouped into three SU(2) singlets, one triplet and one quintet.
The reduced tensor $\mathcal{W}_H$ is therefore labeled by the degeneracy set $d=\{(0,3),(1,1),(2,1)\}$. One then finds for the first site $i=1$ the $\mathcal{W}_H$ matrices,
\begin{eqnarray}
\mathcal{W}_H^{[1],0}&=&
%\sqrt{2s+1}
\begin{pmatrix} 0 & w_0 & 1 & 0 & 0 \\ 0 & 0 & w_0 \cdot \Sigma_0 & 0 & 0 \\ 0 & 0 & 0 & 0 & 0 \\ 0 & 0 & 0 & 0 & 0 \\ 0 & 0 & 0 & 0 & 0 \end{pmatrix}, \nonumber\\
\mathcal{W}_H^{[1],1}&=&
%\sqrt{2s+1}
\begin{pmatrix} 0 & 0 & 0 & w_1 & 0 \\ 0 & 0 & 0 & 0 & 0 \\ 0 & 0 & 0 & 0 & 0 \\ 0 & 0 & w_1 \cdot \Sigma_1 & 0 & 0 \\ 0 & 0 & 0 & 0 & 0 \end{pmatrix}, \\
\mathcal{W}_H^{[1],2}&=&
%\sqrt{2s+1}
\begin{pmatrix} 0 & 0 & 0 & 0 & w_2 \\ 0 & 0 & 0 & 0 & 0 \\ 0 & 0 & 0 & 0 & 0 \\ 0 & 0 & 0 & 0 & 0 \\ 0 & 0 & w_2 \cdot \Sigma_0 & 0 & 0 \end{pmatrix}.\nonumber
\end{eqnarray}
And for all other sites $i=2, \cdots, N$ one obtains
\begin{eqnarray}
\mathcal{W}_H^{[i],0}&=&
%\sqrt{2s+1}
\begin{pmatrix} 1 & 0 & 0 & 0 & 0 \\ w_0 \cdot \Sigma_0 & 0 & 0 & 0 & 0 \\ 0 & w_0 & 1 & 0 & 0 \\ 0 & 0 & 0 & 0 & 0 \\ 0 & 0 & 0 & 0 & 0 \end{pmatrix},\nonumber \\
\mathcal{W}_H^{[i],1}&=&
%\sqrt{2s+1}
\begin{pmatrix} 0 & 0 & 0 & 0 & 0 \\ 0 & 0 & 0 & 0 & 0 \\ 0 & 0 & 0 & w_1 & 0 \\ w_1 \cdot \Sigma_1 & 0 & 0 & 0 & 0 \\ 0 & 0 & 0 & 0 & 0 \end{pmatrix}, \\
\mathcal{W}_H^{[i],2}&=&
%\sqrt{2s+1}
\begin{pmatrix} 0 & 0 & 0 & 0 & 0 \\ 0 & 0 & 0 & 0 & 0 \\ 0 & 0 & 0 & 0 & w_2 \\ 0 & 0 & 0 & 0 & 0 \\ w_2 \cdot \Sigma_0 & 0 & 0 & 0 & 0 \end{pmatrix}\nonumber,
\end{eqnarray}
with %$\omega_0=1$,
\begin{eqnarray}
w_0&=&\sqrt{\frac{1}{3} \, |\sin\theta| \, s^2(s+1)^2},\nonumber\\
w_1&=&\sqrt{\frac{3}{2} \, |\sin\theta-2\cos\theta| \, s(s+1)},\\
w_2&=&\sqrt{\frac{10}{3} \, |\sin\theta| \, (s-\frac{1}{2})s(s+1)(s+\frac{3}{2})}\nonumber
\end{eqnarray}
and $\Sigma_0={\rm Sgn}(\sin\theta)$ and $\Sigma_1={\rm Sgn}(\sin\theta-2\cos\theta)$.
It follows that for $s=1/2$ we have $w_2=0$ and $\mathcal{W}^2=0$. Due to PBC the tensors at the first site have a different structure than the tensors at the other sites.

The reduced tensors of the corresponding unity MPOs are given by
\begin{equation}
\mathcal{W}_1^{[i],0}= {\rm diag} \{\sqrt{2j+1}\},
\end{equation}
i.e. a diagonal matrix constructed from all $j$'s of the degeneracy set $d$.
E.g. for the set $d=\{(0,3),(1,1),(2,1)\}$ the reduced unity operator is given by
\begin{equation}
\mathcal{W}_1^{0}=
\begin{pmatrix} 1 & 0 & 0 & 0 & 0 \\ 0 & 1 & 0 & 0 & 0 \\ 0 & 0 & 1 & 0 & 0 \\ 0 & 0 & 0 & \sqrt{3} & 0 \\ 0 & 0 & 0 & 0 & \sqrt{5} \end{pmatrix}.
\end{equation}
This matrix is used at the site of the fictitious spin.

For the Heisenberg antiferromagnet ($\theta=0$) or ferromagnet ($\theta=\pi$) the MPOs can be simplified. One obtains: $w_0=w_2=0$, $\Sigma_0=0$, $\Sigma_1=-{\rm Sgn}(\cos\theta)=\mp 1$, and $\mathcal{W}^2=0$. Therefore, the quintet and one singlet in the basis can be omitted, so that the basis is $\{(0,2),(1,1)\}$. Then, the reduced tensors take the form
for $i=1$
\begin{equation}
\mathcal{W}_H^{[1],0}=
%\sqrt{2s+1}
\begin{pmatrix} 0 & 1 & 0 \\ 0 & 0 & 0 \\ 0 & 0 & 0 \end{pmatrix}, \hspace{0.3 cm} \mathcal{W}_H^{[1],1}=
%\sqrt{2s+1}
\begin{pmatrix} 0 & 0 & \omega_1 \\ 0 & 0 & 0 \\ 0 & \omega_1 \cdot \Sigma_1 & 0 \end{pmatrix},
\end{equation}
and for the other sites $i=2,\dots,N$
\begin{equation}
\mathcal{W}_H^{[i],0}=
%\sqrt{2s+1}
\begin{pmatrix} 1 & 0 & 0 \\ 0 & 1 & 0 \\ 0 & 0 & 0 \end{pmatrix}, \hspace{0.3 cm}
\mathcal{W}_H^{[i],1}=
%\sqrt{2s+1}
\begin{pmatrix} 0 & 0 & 0 \\ 0 & 0 & \omega_1 \\ \omega_1 \cdot \Sigma_1 & 0 & 0 \end{pmatrix}
\end{equation}
with $\omega_1=\sqrt{3s(s+1)}$.
The reduced unity tensor at the site of the fictitious spin shrinks to
\begin{equation}
\mathcal{W}_1^{0}=
\begin{pmatrix} 1 & 0 & 0 \\ 0 & 1 & 0 \\ 0 & 0 & \sqrt{3} \end{pmatrix}.
\end{equation}

\begin{widetext}
\section{Reduced block tensors}\label{reducedTensors}

The calculation of matrix elements of MPS in MPO Eq.~(\ref{eq:matrixe}) reduces to a finite product of transfer tensors (`blocks').
Similarly, the block tensors
$H_L$, $H_R$, $N_L$, $N_R$, $O_{L,k}$, $O_{R,k}$ which are necessary to build the generalized eigenvalue
problem~(\ref{eq:eff-Hamiltonian-p}), are finite products of transfer operators. The size of such blocks may be increased by multiplication of a transfer tensor on the right or on the left of an existing block. Generically, we will denote these block tensors by $B_L$ and $B_R$, and the index distinguishes if the block
is produced by left or the right multiplication.

In reduced form these tensors have three intermediate indices $e_1$, $e_2$, $e_3$, and, therefore, a reduced block tensor $\mathcal{B}_L$ has rank 9. To define these reduced tensors we choose the coupling scheme shown in Fig.~\ref{fig:fstreeblock}, which corresponds to the following expression
\begin{multline}\label{redblocktensor}
(B_L)_{(\bar{I},\bar{a},\bar{b}),(I,a,b)}=
\sum_{\substack{j_{e_1},m_{e_1} \\ j_{e_2},m_{e_2} \\ j_{e_3},m_{e_3}}} (\mathcal{B}_L^{j_{e_1},j_{e_2},j_{e_3}})_{(\gamma_{\bar{I}},\gamma_{\bar{a}},\gamma_{\bar{b}}),(\gamma_I,\gamma_a,\gamma_b)} C_{m_{\bar{a}},m_{e_1},m_{\bar{b}}}^{j_{\bar{a}},j_{e_1},j_{\bar{b}}} \,
C_{m_{\bar{I}},{m_{e_3}},m_{e_1}}^{j_{\bar{I}},j_{e_3},j_{e_1}} \, C_{m_a,m_{e_2},m_b}^{j_a,j_{e_2},j_b} \, C_{m_I,m_{e_3},m_{e_2}}^{j_I,j_{e_3},j_{e_2}}
\end{multline}
We can take $t_{e_1}=t_{e_2}=t_{e_3}=1$ without loss of generality.
\begin{figure}
\unitlength1cm
\begin{picture}(6,3)(0,0)
 \put(2.,0)  {\includegraphics[width=2.5 cm]{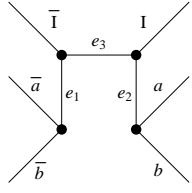}}
\end{picture}
\caption{\footnotesize Coupling scheme corresponding to the decomposition of a block tensor $H$, $N$ or $O_k$ (see Eq.~(\ref{redblocktensor})). Each vertex represents a C factor or 3$j$ symbol. Summation over the indices of internal edges is implied.\label{fig:fstreeblock}}
\end{figure}
For $N$- and $O$-blocks  $j_{\bar{I}}=j_{I}=0$ (and $\gamma_{\bar{I}}=\gamma_{I}=1$) and, consequently,  $j_{e_1}=j_{e_2}=j_{e_3}$. For right blocks a similar expression may be written down. Using the orthogonality relations of the $3j$ symbols~\cite{edmonds}, one can
express the reduced tensor elements of the reduced block $\mathcal{B}_L$ in terms of the tensor elements of $B_L$.

During the update procedure of the algorithm the length of the blocks has to be increased by right or left multiplication of transfer tensors as
illustrated in Fig.~\ref{fig:tm}. However, we only need to determine reduced blocks and want to avoid reconstruction of the full block. In order to do that the summation over intermediate spin projection indices must be carried out
analytically using appropriate summation formulas for $3j$ symbols. Such formulas can be found e.g. in Edmonds~\cite{edmonds}, and we will strictly follow the conventions of that reference.

An increase of the reduced left block by one transfer matrix may be formulated in terms of the following recursion formula,
\begin{multline}\label{blockrecursion}
(\mathcal{B}_L^{[i+1],j_{e_1},j_{e_2},j_{e_3}})_{(\gamma_{\bar{I}},\gamma_{\bar{a}},\gamma_{\bar{b}}),
(\gamma_I,\gamma_a,\gamma_b)} = (2j_{e_2}+1) \, \sum_{\substack{j_{e},j_{\underline{e}_2} \\ {\gamma_{\underline{I}},\gamma_{\underline{a}},\gamma_{\underline{b}}}}} \, \begin{bmatrix} j_b & s_i & j_{\underline{b}} \\ j_{I} & j_{e} & j_{\underline{I}} \\ j_{a} & s_i & j_{\underline{a}} \\ j_{e_2} & j_{e_3} & j_{\underline{e}_2} \end{bmatrix} \, (\mathcal{B}_L^{[i],j_{e_1},j_{\underline{e}_2},j_{e_3}})_{(\gamma_{\bar{I}},\gamma_{\bar{a}},\gamma_{\bar{b}}),
(\gamma_{\underline{I}},\gamma_{\underline{a}},\gamma_{\underline{b}})} \times \\
\times \sqrt{2s_i+1} \,\, \mathcal{W}_{\gamma_{\underline{I}},\gamma_I}^{[i],j_{e}} \, \alpha_{\gamma_{\underline{a}},\gamma_{a}}^{[i]*} \, \mathcal{\beta}_{\gamma_{\underline{b}},\gamma_{b}}^{[i]}.
\end{multline}
The analogous formula for right block is
\begin{multline}\label{blockrecursion}
(\mathcal{B}_R^{[i-1],j_{e_1},j_{e_2},j_{e_3}})_{(\gamma_{I^{\prime}},\gamma_{a^{\prime}},\gamma_{b^{\prime}}),
(\gamma_{\bar{I}},\gamma_{\bar{a}},\gamma_{\bar{b}})} = (2j_{e_1}+1) \, \sum_{\substack{j_{e},j_{\underline{e}_1} \\ {\gamma_{\underline{I}},\gamma_{\underline{a}},\gamma_{\underline{b}}}}} \, \begin{bmatrix} j_{\underline{b}} & s_i & j_{b^{\prime}} \\ j_{\underline{I}} & j_{e} & j_{I^{\prime}} \\ j_{\underline{a}} & s_i & j_{a^{\prime}} \\ j_{\underline{e}_1} & j_{e_3} & j_{e_1} \end{bmatrix} \, (\mathcal{B}_R^{[i],j_{\underline{e}_1},j_{e_2},j_{e_3}})_{(\gamma_{\underline{I}},\gamma_{\underline{a}},
\gamma_{\underline{b}}),(\gamma_{\bar{I}},\gamma_{\bar{a}},\gamma_{\bar{b}})} \times \\
\times \sqrt{2s_i+1} \,\, \mathcal{W}_{\gamma_{I^{\prime}},\gamma_{\underline{I}}}^{[i],j_{e}} \, \alpha_{\gamma_{a^{\prime}},\gamma_{\underline{a}}}^{[i]*} \, \mathcal{\beta}_{\gamma_{b^{\prime}},\gamma_{\underline{b}}}^{[i]}.
\end{multline}
Here
\begin{equation}
\begin{bmatrix} j_{11} & j_{12} & j_{13} \\ j_{21} & j_{22} & j_{23} \\ j_{31} & j_{32} & j_{33} \\ j_{41} & j_{42} & j_{43} \end{bmatrix}=(-1)^{j_{21}+j_{22}+j_{42}+j_{43}} \, \Biggl\{\begin{matrix} j_{11} & j_{12} & j_{13} \\ j_{41} & j_{22} & j_{43} \\ j_{31} & j_{32} & j_{33} \end{matrix}\Biggr\} \, \biggl\{\begin{matrix} j_{21} & j_{22} & j_{23} \\ j_{43} & j_{42} & j_{41} \end{matrix}\biggr\}.
\end{equation}
The expressions in curly brackets denote the Wigner $9j$ and the Racah $6j$ symbol, respectively. The matrices $\alpha$ and $\beta$ are the reduced tensors corresponding to the MPS tensors $A$ and $B$, respectively, and $\mathcal{W}$ is the reduced $W$ tensor of an MPO. The local spin $s_i=s$ for $i \le N$ and $J$ for the fictitious site. In order to avoid unnecessary indices we again do not label this expression with the local spin index, which is fixed in our calculation.

The recursion is started from a reduced identity block tensor, which is easily obtained from Eq.~(\ref{redblocktensor}):
\begin{equation}
\mathcal{I}_{(\gamma_{\bar{I}},\gamma_{\bar{a}},\gamma_{\bar{b}}), (\gamma_{\underline{I}},\gamma_{\underline{a}},\gamma_{\underline{b}})}^{j_{e_1},j_{e_2},j_{e_3}}= (2j_{e_1}+1) (2j_{e_3}+1) \, \delta_{\gamma_{\bar{I}},\gamma_{\underline{I}}} \, \delta_{\gamma_{\bar{a}},\gamma_{\underline{a}}} \, \delta_{\gamma_{\bar{b}},\gamma_{\underline{b}}} \, \delta_{j_{e_1},j_{e_2}} \, \delta (j_{\bar{a}},j_{e_1},j_{\bar{b}}) \, \delta (j_{\bar{I}},j_{e_3},j_{e_1})
\end{equation}
with $\delta(j_1,j_2,j_3)=1$ if $j_1,j_2,j_3$ fulfill the `triangle rule' and zero otherwise.
Superficially it appears that the recursion formula for  $\mathcal{B}_L$ is (appart from a trivial prefactor) independent of $j_1$ and analogously $\mathcal{B}_R$ independent of $j_2$. However, this is in fact not the case since both $j_1$ and $j_2$ enter via the initial condition of the recursion. We remark that the reduced block tensors $\mathcal{B}$ are {\it very} sparse. Therefore, an implementation of arbitrary rank sparse tensors is needed for their calculation and manipulation.

\section{Optimization step in terms of reduced tensors}\label{redeig}

The reduced effective Hamiltonian $\mathcal{H}_{\rm{eff}}$ and the reduced effective normalization matrix $\mathcal{N}_{\rm{eff}}$ for the generalized eigenvalue problem~(\ref{redupdate}) are obtained from
\begin{multline}\label{Oeff}
(\mathcal{O}_{\rm{eff}}^{[i]})_{[\gamma_{a^\prime} \gamma_a],[\gamma_{b^\prime} \gamma_b]} =  \sum_{\substack{j_{e},j_{e_1},j_{e_2},j_{e_3},j_{e_1^\prime} \\ \gamma_I,\gamma_{I^{\prime}},\gamma_{\bar{I}},\gamma_{\bar{a}},\gamma_{\bar{b}}}} \frac{\sqrt{2s_i+1}}{(2j_{e_1}+1)(2j_{e_3}+1)} \, \begin{bmatrix} j_{b^\prime} & s_i & j_{b} \\ j_{I^{\prime}} & j_{e} & j_I \\ j_{a^{\prime}} & s_i & j_a \\ j_{e_1^\prime} & j_{e_3} & j_{e_2} \end{bmatrix} \, \mathcal{W}_{\gamma_I,\gamma_{I^{\prime}}}^{[i],j_{e}} \, (\mathcal{B}_R^{[i],j_{e_1^{\prime}},j_{e_1},j_{e_3}})_{(\gamma_{I^{\prime}}\gamma_{a^\prime}\gamma_{b^{\prime}}), [\gamma_{\bar{I}}\gamma_{\bar{a}} \gamma_{\bar{b}}]} \times \\
\times (\mathcal{B}_L^{[i],j_{e_1},j_{e_2},j_{e_3}})_{[\gamma_{\bar{I}}\gamma_{\bar{a}} \gamma_{\bar{b}}],(\gamma_I\gamma_a \gamma_b)}
\end{multline}
Of course, for $\mathcal{N}_{\rm{eff}}$ one has to take the    $\mathcal{W}_{\gamma_{I},\gamma_{I^{\prime}}}^{[i],j_{e}}$ of the identity operator, which leads
to important simplifications in the $6j$ and $9j$ symbols.

In the same way one also obtains the reduced vector $y_k$ for the calculation of excited states according to Eq.~(\ref{projector}), where the index $k$ labels the different excited states,
\begin{multline}
(y_k^{[i]})_{[\gamma_{a^{\prime}} \gamma_a]}=\sum_{\substack{j_{e_1}\\ \gamma_b,\gamma_{b^{\prime}},\gamma_{\bar{a}},\gamma_{\bar{b}}}} \, \frac{1}{(2j_{e_1}+1)^2} \, \begin{bmatrix} j_{b^\prime} & s_i & j_{b} \\ 0 & 0 & 0 \\ j_{a^{\prime}} & s_i & j_a \\ j_{e_1} & j_{e_1} & j_{e_1} \end{bmatrix} \, \sqrt{2s_i+1} \,\, (\mathcal{B}_{R,k}^{[i],j_{e_1},j_{e_1},j_{e_1}})_{(1,\gamma_{a^{\prime}}, \gamma_{b^{\prime}}),[1 \, \gamma_{\bar{a}} \, \gamma_{\bar{b}}]} \times \\
\times (\mathcal{B}_{L,k}^{[i],j_{e_1},j_{e_1},j_{e_1}})_{[1 \,\gamma_{\bar{a}} \, \gamma_{\bar{b}}],(1,\gamma_a,\gamma_b)} \, (\mathcal{\phi}_k^{[i]})_{\gamma_b,\gamma_{b^{\prime}}}
\end{multline}
where $\mathcal{\phi}_k^{[i]}$ denotes the reduced tensor of the MPS tensor $\Phi_k^{[i]}$ in Eq.~(\ref{eq:proj1}). Analogously, the block tensors $\mathcal{B}$ that correspond here to $O$ tensors are labeled by the additional index $k$.

\end{widetext}

\section{Calculation of $\langle H^2 \rangle$}\label{Hsquare}

The MPO of a squared operator is needed for the calculation of the expectation value $\langle H^2 \rangle$. It is constructed as explained in Appendix B of Ref.~\cite{PhysRevB.93.054417}. The two left and the two right virtual indices are fused into a single virtual index each, and a summation over intermediate spin projections of the physical spin $s$ is performed.

In the SU(2) symmetric formalism, we need to express the reduced $\mathcal{W}$ tensor of the squared operator $O^2$ in terms of the reduced $\mathcal{W}$ tensor of the operator $O$. To this end we follow section 7.1 of Edmonds~\cite{edmonds}. Using Eq.~(7.1.1) and Eq.~(7.1.5) of this reference one obtains
\begin{eqnarray}\label{wsquare}
& &(\mathcal{W}_{O^2})_{[(j_1 t_1) (j_2 t_2)](j t),~[(j_1^{\prime} t_1^{\prime}) (j_2^{\prime} t_2^{\prime})](j^{\prime} t^{\prime})}^{j_e}=\\
& &~~~
%(-1)^{j_e+2s}
= (2j_e+1)\sqrt{(2j+1)(2j^{\prime}+1)(2s+1)} \, (-1)^{j_{e_2}} \times \nonumber\\
& &~~~\times
%\Biggl\{\begin{matrix} j_1 & j_1^{\prime} & j_{e_1} \\ j_2 & j_2^{\prime} & j_{e_2} \\ j & j^{\prime} & j_e \end{matrix}\Biggr\} \, \biggl\{\begin{matrix} j_{e_1} & j_{e_2} & j_e \\ s & s & s \end{matrix}\biggr\}
\begin{bmatrix} j_1 & j & j_2 \\ s & j_e & s \\ j_1^{\prime} & j^{\prime} & j_2^{\prime} \\ j_{e_1} & s & j_{e_2} \end{bmatrix}
%\,  \nonumber\\ & &~~~\times
\, (\mathcal{W}_O)_{(j_1 t_1),(j_1^{\prime} t_1^{\prime})}^{j_{e_1}} (\mathcal{W}_O)_{(j_2 t_2),(j_2^{\prime} t_2^{\prime})}^{j_{e_2}}\nonumber
\end{eqnarray}
Here, $j_{e_1}$ and $j_{e_2}$ are the intermediate indices of reduced $\mathcal{W}_{O}$ tensors while $j_e$ is the intermediate index of reduced $\mathcal{W}_{O^2}$ tensor. The left virtual indices $(j_1 t_1)$ and $(j_2 t_2)$ of $\mathcal{W}_O$ tensors are fused into one index $(j t)$, and, analogously, the right indices $(j_1^{\prime} t_1^{\prime})$ and $(j_2^{\prime} t_2^{\prime})$ are fused into the index $(j^{\prime} t^{\prime})$.
%In the case of the MPOs the sums in Eq.~(7.1.1) and (7.1.5) of~\cite{edmonds} appear to consist of only one term.
One can immediately read off Eq.~(\ref{wsquare}) that for the bilinear-biquadratic spin-1 Heisenberg model the reduced $\mathcal{W}_{H^2}$ tensors have dimensions $3 \times 36 \times 36$ with the virtual spins described by the degeneracy set $\{(0,11),(1,11),(2,10),(3,3),(4,1)\}$. These tensors are rather sparse. Note, that the result (\ref{wsquare}) holds for any spin $s$.

\section{Regauging SU(2)-symmetric MPS}\label{regauging}

%The expresions $Q^L$ and $Q^R$ defined in Eq.~(\ref{LRnormalization}) can be brought into reduced form.
Using Eq.~(\ref{mel}) and the orthogonality relations of the $3j$ symbols it is easy to show that
the expressions $Q^L$ and $Q^R$ defined in  Eq.~(\ref{LRnormalization}) are block diagonal. For instance, the matrix element of $Q^R$

\begin{eqnarray}
&&Q_{(j^{\prime},t^{\prime},m^{\prime}),(j^{\prime \prime},t^{\prime \prime},m^{\prime \prime})}^R=
 \\
&=&\delta_{j^{\prime},j^{\prime \prime}} \delta_{m^{\prime},m^{\prime \prime}} \cdot (\frac{1}{2j^{\prime}+1} \sum_{j,t} \mathcal{M}^s_{\gamma^{\prime},\gamma} (\mathcal{M}^{s\dagger})_{\gamma,\gamma^{\prime \prime}} \delta(j,s,j^{\prime}))\nonumber\\
&=&\delta_{j^{\prime},j^{\prime \prime}} \delta_{m^{\prime},m^{\prime \prime}} \tilde{Q}_{\gamma^{\prime},\gamma^{\prime \prime}}^R\nonumber.
\end{eqnarray}
An analogous result can be obtained for $Q^L$. Therefore, the reduced tensor $\tilde{Q}^{R/L}$ can be diagonalized blockwise. We then define the regauged reduced MPS $\tilde{\mathcal{M}}^s=(\tilde{Q}^{R})^{-1/2} \mathcal{M}^s$ or $\tilde{\mathcal{M}}^s=\mathcal{M}^s (\tilde{Q}^{L})^{-1/2}$ (if $\tilde{Q}^{L/R}$ cannot be inverted, the pseudoinverse is used instead). It can be easily shown that these regauged
MPSs fulfill the left/right normalization conditions  Eq.~(\ref{LRnormalization}), respectively.

\section{Performance of the algorithm}\label{spin-1/2}

In order to test the performance of the proposed algorithm and its implementation we present a few results for the bi-linear Heisenberg model ($\theta=0$ in Eq.~(\ref{eq-bilbiq})) for periodic spin-1/2 systems.
Benchmark results for comparison  are obtained from the finite-size Bethe Ansatz~\cite{Karbach1998}.

In Table~\ref{dataHeisenberg12} we present ground state energies for systems of up to 50 sites with total spin $J=0$ and $J=1$. The Bethe Ansatz results are reproduced with high precision. The (rather moderate) computational effort depends on the degeneracy sets chosen
for the virtual spins of the MPSs. Our choice is listed in Table~\ref{dataHeisenberg12}. Not surprisingly,
for larger systems we need bigger degeneracies for each virtual spin.  As mentioned already in the main text, at this stage the virtual spins and their degeneracies have to be pre\-selected, and they are kept fixed throughout the calculation.

\begin{table*}
\begin{tabular}{|c|c|c|c|c|c||c|c|c|c|c|}
  \hline
  $N$ & $J$ & $E/N$ & $E_{\rm BA}/N$ & $\Delta E/E_{\rm BA}$ & $d$ & $J$ & $E/N$ & $E_{\rm BA}/N$ & $\Delta E/E_{\rm BA}$ & $d$ \\
  \hline
  10 & 0 & -0.45154463545 & -0.45154463545 & $ < 10^{-11}$       & $\{3, 3, 3, 3, 3\}$ &
       1 & -0.40922073467 & -0.40922073467 & $ < 10^{-11}$       & $\{4, 4, 4, 4, 4\}$ \\
  \hline
  20 & 0 & -0.44521898185 & -0.44521932640 & $7.7 \cdot 10^{-7}$ & $\{4, 4, 4, 4, 4\}$ &
       1 & -0.43432031232 & -0.43432204931 & $4.0 \cdot 10^{-7}$ & $\{5, 5, 5, 5, 5\}$ \\
  \hline
  30 & 0 & -0.44406518996 & -0.44406543530 & $5.5 \cdot 10^{-7}$ & $\{4, 4, 4, 4, 4\}$ &
       1 & -0.43915821100 & -0.43916046223 & $5.1 \cdot 10^{-6}$ & $\{5, 5, 5, 5, 5\}$ \\
  \hline
  40 & 0 & -0.44366290056 & -0.44366306970 & $3.8 \cdot 10^{-7}$ & $\{6, 6, 6, 6, 6\}$ &
       1 & -0.44088201219 & -0.44088325383 & $2.8 \cdot 10^{-6}$ & $\{6, 6, 6, 6, 6\}$  \\
  \hline
  50 & 0 & -0.44347688297 & -0.44347713230 & $5.6 \cdot 10^{-7}$ & $\{6, 6, 6, 6, 6\}$ &
       1 & -0.44168120000 & -0.44168889567 & $1.7 \cdot 10^{-6}$ & $\{6, 6, 6, 6, 6\}$\\
  \hline
\end{tabular}
\caption{\footnotesize Ground state energy per site $E/N$ in the $J=0$ and $J=1$ sectors for the spin-1/2 isotropic Heisenberg ring as a function of $N$. $E_{\rm BA}$ are results from finite size Bethe Ansatz~\cite{Karbach1998}.
The degeneracy sets are $d=\{(0,x_1),(1/2,x_2),(1,x_3),(3/2,x_4),(2,x_5)\}$ which we abbreviate as $\{x_1,x_2,x_3,x_4,x_5\}$. \label{dataHeisenberg12}}
\end{table*}

We briefly analyze the performance of the algorithm as a function of the degeneracy $t$ of the symmetry sectors for a ring with 30 spins, and determine the memory requirements (RAM), CPU time as well the relative error $\Delta E/E_{BA}$ of the $J=0$ state as a function of $t$. We use the degeneracy set $d=\{(0,t),(1/2,t),(1,t),(3/2,t),(2,t)\}$. Since the calculations with different $t$ may take different number of sweeps to satisfy the prescribed convergence criteria, we will calculate the CPU time required for the optimization of the MPS at one site.

RAM scales with $t$ roughly as $O(t^{4})$ in line with the fact that `block' tensors have 4 indices corresponding to MPS virtual indices.
CPU time per site is proportinal to $t$, i.e. $O(m)$ for the spin 1/2 model. However for the BBH Hamiltonian we observe a scaling of about $O(t^5)$.
The precision of the result measured by the relative error $\Delta E/E_{\rm{BA}}$ improves roughly exponentially slow with
increasing $t$.

The performance of the algorithm as a function of $j_{\rm max}$ for the virtual spins is difficult to assess
at this stage as we can at present only use $j_{\rm max}<4$. An algorithmic selection of $j_{\rm max}$ with appropriate
degeneracy $t_{\rm max}$ for each $j$ is desirable, and an implementation of such a procedure is currently underway.

\section{Energy of the states with $J=N$ in spin-1 models}\label{marginalspin}

For this case  the trivial virtual spin representation $d=\{(N/2,1)\}$ is sufficient. Therefore, the {\it reduced} local MPS tensor will be $1 \times 1$ matrix, as will also be the matrices $\mathcal{N}_{\rm eff}$ and $\mathcal{H}_{\rm eff}$.

According to translational invariance and the regauging rules (see Appendix~\ref{regauging}), the reduced MPS tensor at each site (except the `fictitious site') will be the same and equal to $\mathcal{M}=(\sqrt{N+1})$. Therefore, one can easily evaluate $\mathcal{N}_{\rm eff}$, as it is independent of the model under investigation. In this case the intermediate indices are always equal, therefore
\begin{multline}
(\mathcal{N}_{\rm eff})_{(1,1),(1,1)}=\sum_{j_{e_1}} \, (-1)^{j_{e_1}} \, (2j_{e_1}+1) \times \\
\times \frac{(N!)^2}{(N-j_{e_1})!(N+j_{e_1}+1)!} \, \left(1-\frac{2j_{e_1}(j_{e_1}+1)}{N(N+2)}\right)^N.
\end{multline}

The matrix $\mathcal{H}_{\rm eff}$ depends on the Hamiltonian under consideration. For the spin-1 bilinear-biquadratic Heisenberg model one finds
\begin{multline}
(\mathcal{H}_{\rm eff})_{(1,1),(1,1)}=N(\cos \theta + \sin \theta) \, \sum_{j_{e_1}} \, (-1)^{j_{e_1}} \, (2j_{e_1}+1) \times \\
\times \frac{(N!)^2}{(N-j_{e_1})!(N+j_{e_1}+1)!} \, \left(1-\frac{2j_{e_1}(j_{e_1}+1)}{N(N+2)}\right)^N.
\end{multline}
Consequently, one obtains the energy of the $J=N$ state as $E=\frac{1}{N}\frac{(\mathcal{H}_{\rm eff})_{(1,1),(1,1)}}{(\mathcal{N}_{\rm eff})_{(1,1),(1,1)}}=\cos\theta+\sin\theta$ as expected.
Despite the analytical simplicity of this result it cannot be obtained numerically within the present implementation. As is obvious from the analytical results the matrix elements of matrices $\mathcal{H}_{\rm eff}$ and $\mathcal{N}_{\rm eff}$ decrease with increasing $N$. At about $N \simeq 25$ the numerical values cannot be represented by machine precision numbers any more, and the calculated values are arbitrary.

\end{document}